\documentclass[graybox]{svmult}

\usepackage{mathptmx}       
\usepackage{helvet}         
\usepackage{courier}        
\usepackage{type1cm}        
\usepackage{makeidx}         
\usepackage{graphicx}        
\usepackage{multicol}        
\usepackage[bottom]{footmisc}

\makeindex             

\begin{document}

\title*{Black Holes at the LHC}
\author{Panagiota Kanti}
\institute{Panagiota Kanti \at Division of Theoretical Physics, Physics Department,
University of Ioannina, Ioannina GR-45110, Greece. \email{pkanti@cc.uoi.gr}}
%
%
\maketitle

\abstract{In these two lectures, we will address the topic of the creation of
small black holes during particle collisions in a ground-based accelerator,
such as LHC, in the context of a higher-dimensional theory. We will cover
the main assumptions, criteria and estimates for their creation, and we will
discuss their properties after their formation. The most important observable
effect associated with their creation is likely to be the emission of Hawking
radiation during their evaporation process. After presenting the mathematical
formalism for its study, we will review the current results for the emission
of particles both on the brane and in the bulk. We will finish with a 
discussion of the methodology that will be used to study these spectra,
and the observable signatures that will help us identify the black-hole events.} 

\section{Introduction}
\label{kanintro}

These two lectures aim at offering an introduction to the idea that miniature
black holes may be created during high-energy particle collisions at ground-based
colliders. This scenario can only be realised in the context of higher-dimensional
theories, i.e. theories that postulate the existence of additional spacelike
dimensions in nature. An introduction to the two most important versions of these
theories, namely the scenario with Large Extra Dimensions and the one with Warped
Extra Dimensions will be our starting point.    

We will then proceed to introduce the idea of the possible creation of black
holes at the laboratory. We will present some simple but illuminating geometrical
criteria for this to happen. We will then discuss the boundary value problem whose
solution determines whether a black hole has been formed out of two colliding
particles. Certain aspects of the creation process will be studied in more detail,
namely the amount of energy that is absorbed by the created black hole and the
value of the production cross-section. We will finally discuss the properties
of the produced black holes, such as the horizon value, temperature and lifetime,
and compare with the ones of their 4-dimensional analogues. The non-vanishing,
in general, temperature of the black hole is associated with the emission of
a thermal type of radiation from the black hole, i.e. the Hawking radiation.
This has its source at the creation of a virtual pair of particles outside
the horizon of the black hole (or, equivalently, the quantum tunneling of a
particle from within the black hole horizon). We will finish our first lecture
with a brief outline of the mathematical formalism that was developed for
the study of the Hawking radiation. 

The emission of Hawking radiation, i.e. of elementary particles with a thermal
spectrum, takes place during the two intermediate phases in the life of a black
hole. These are the spin-down phase and the Schwarzschild phase, in chronological
order. Starting from the second, that has the simplest gravitational background,
we will present a review of the results that have been derived in the literature
related to the form of the radiation spectra and their most characteristic
features, including their dependence on the dimensionality of spacetime and
the relative emissivities of different species of fields. A similar task will
then be taken for the spin-down phase during which the black hole carries a
non-vanishing angular momentum. In this case, the radiation spectra will have
an extra dependence on the angular momentum parameter of the black hole, as
well as an angular distribution in space due to the existence of a preferred
direction in space, that of the rotation axis. The most important part, from the
phenomenological point of view, will be the emission of the black hole directly
on the brane on which the Standard Model particles and the observers themselves
are located. However, the bulk emission will also be considered as this will
determine the amount of energy remaining for emission on the brane. 

Having completed the theoretical study of the radiation spectra from a
higher-dimensional black hole, we now need to address the question of what
information we may deduce from these spectra, if one day we manage to detect
them, and in what way. Certain properties of the produced black hole such
as the mass and temperature need to be determined first. From these, one
may then turn to the derivation of more fundamental parameters such as
the dimensionality of the gravitational background, or even the value
of the fundamental Planck scale and the cosmological constant. As we
will see, this task is highly non-trivial and demands the close cooperation
of theoretical studies and experimental skill. But if it works, it might
provide answers to the most fundamental questions in theoretical physics. 

%
%

\section{First Lecture: Creation of Black Holes and their Properties}
\label{kanlecture1}

During the first lecture, we will set the stage for the production and
subsequent detection of higher-dimensional black holes. After a brief
introduction to models with extra dimensions, we will discuss the possibility
of the creation of a black hole during a particle collision, and address
certain questions related to this phenomenon. We will then turn to the
evaporation process of the black hole, and we will briefly present
the mathematical formalism for the study of the Hawking radiation.

\subsection{Extra Dimensions}
\label{kanextra}

It is an amazing feature of the Theory of General Relativity that it can be
straightforwardly extended to an arbitrary number of dimensions. Its main
mathematical construction, Einstein's field equations
\begin{equation}
G_{\mu\nu} = R_{\mu\nu} - \frac{1}{2}\,g_{\mu\nu}\,R=
\kappa^2\, T_{\mu\nu}\,,
\label{kanEinstein}
\end{equation}
are expressed in terms of second-rank tensors whose indices can take any
values, depending on our assumptions for the dimensionality of spacetime, without
its mathematical consistency to be in any danger. It comes therefore as
no surprise that, only a few years after Einstein formulated his theory
of gravity, Kaluza produced a gravitational model in five dimensions. The
model was soon supplemented with further suggestions about the topology
of the extra dimension by Klein, and it was the first attempt ever to derive
a unification theory in which gravity played the fundamental role.   

Klein pictured the extra spacelike dimension introduced by Kaluza as a
regular, compact one with finite size ${\cal R}$. To avoid any conflicts
with observational data, the size of the extra dimension was assumed to
be much smaller than any observable length scale. The idea was extensively
used decades later in the formulation of String Theory: there, the size of
the additional six spacelike dimensions, necessary for the mathematical
and physical consistency of the theory, was assumed to be 
${\cal R}=l_{P}=10^{-33}$ {\rm cm}. However, all traditional ideas about
the structure, size and use of the extra space radically changed in the 90's.
The start was made in the context of string theory, where the idea
\cite{kanAntoniadis, kanHW, kanLykken} that the string scale does not
necessarily need to be tied to the Planck scale, $M_P \simeq 10^{19}$ GeV,
was put forward. This soon led\footnote{For some early attempts to
construct higher-dimensional gravitational models, see Refs. \cite{kanAkama,
kanRuS, kanVisser, kanWiltshire}.} to the construction of two, much simpler but
extremely rich from the phenomenological point of view, gravitational models:
the scenario with Large Extra Dimensions \cite{kanADD, kanAADD} and the one
with Warped Extra Dimensions \cite{kanRS}. 

\begin{figure}[t]
\mbox{
\includegraphics[scale=.35]{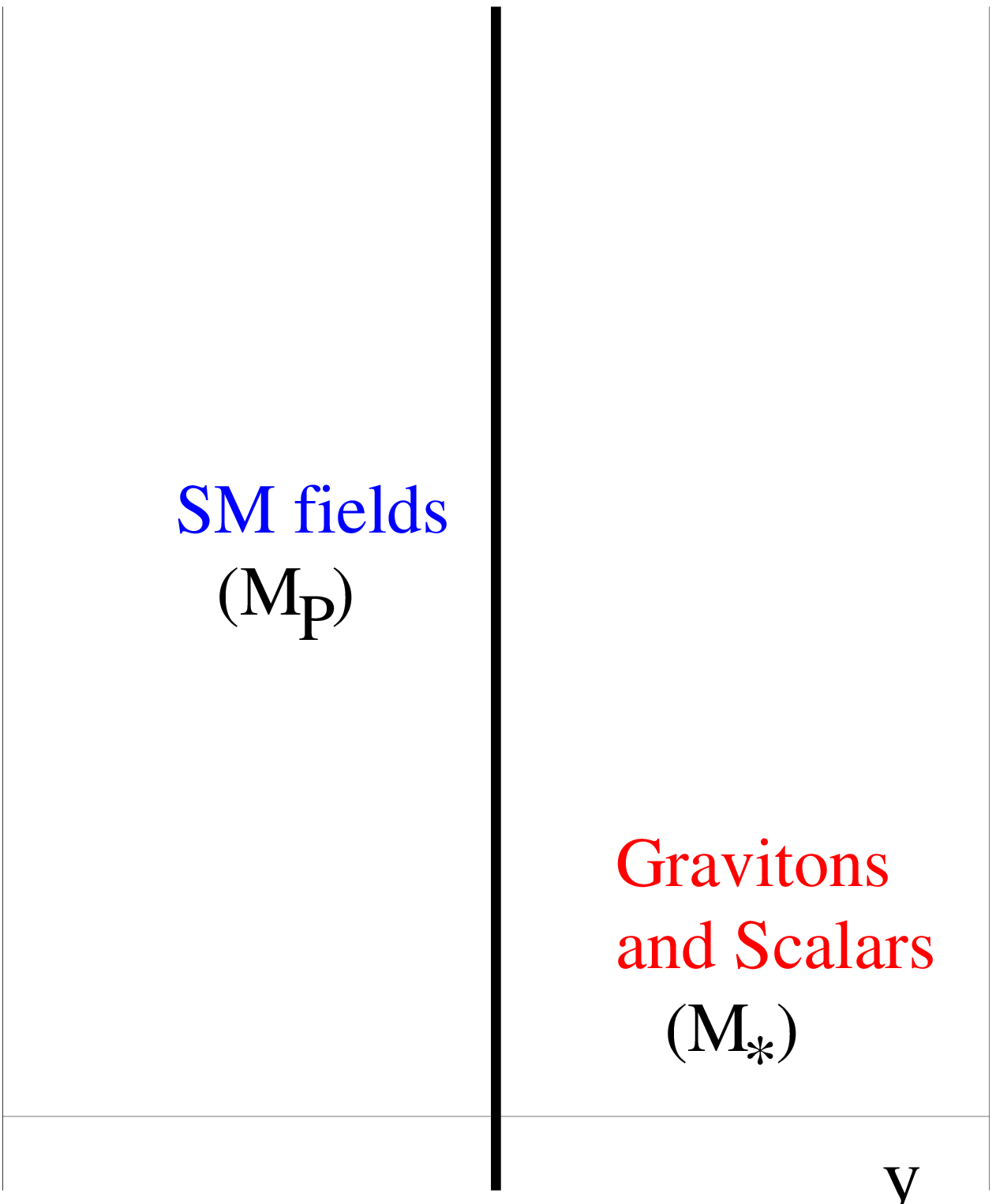}}
\hspace*{1.0cm}
{\includegraphics[scale=.45]{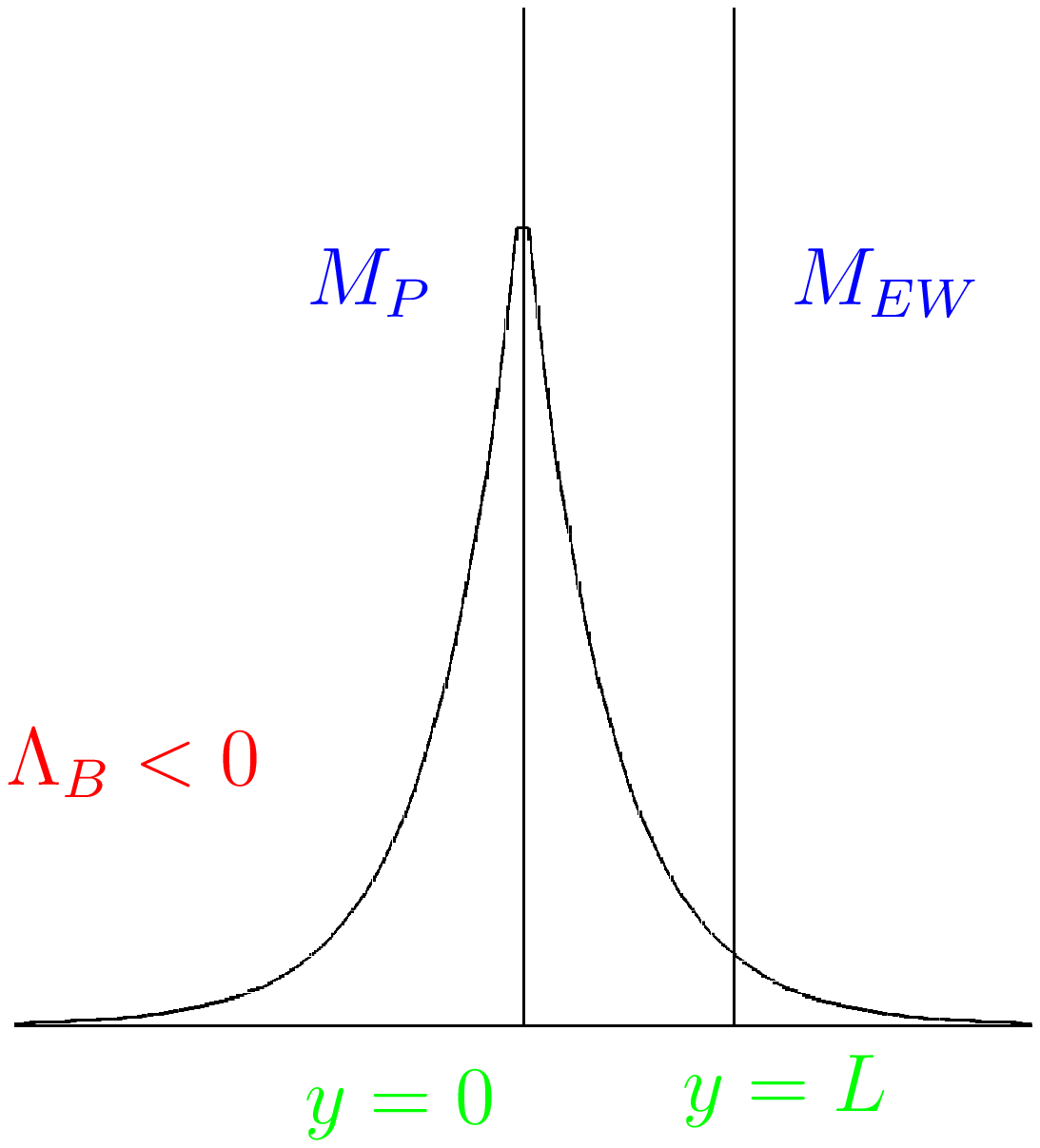}}
\begin{tabular}{cc} \hspace*{2.0cm}{\bf (a)} \hspace*{3.0cm} & \hspace*{3.0cm}
{\bf (b)} \end{tabular}
\caption{{\bf (a)} A 3-brane embedded in a $(4+n)$-dimensional flat spacetime.
{\bf (b)} Two 3-branes embedded in a 5-dimensional Anti de Sitter spacetime.}
\label{kanbranes}
\end{figure}

The topological structure of the higher-dimensional spacetime in each case
is shown in Figs. \ref{kanbranes}(a,b). In the scenario with Large Extra
Dimensions, depicted in Fig. \ref{kanbranes}(a), a 4-dimensional brane is
embedded in a $(4+n)$-dimensional flat space with (3+1) non-compact and
$n$ spacelike compact dimensions. All ordinary matter, made up of and
interacting through Standard Model (SM) fields, is localised on the brane, and
experiences gravitational forces that become strong at Planck scale. On
the other hand, gravitons, and possibly scalars or other fields not carrying any
charges under the SM gauge group, can propagate in the full spacetime. The 
higher-dimensional, fundamental theory has a new scale for gravity, $M_{*}$,
that is related to the effective 4-dimensional one through the equation 
\cite{kanADD, kanAADD}
\begin{equation}
M_P^2 \simeq {\cal R}^{\,n} \,M_{*}^{2+n}\,.
\label{kanM*}
\end{equation}
According to the above, if ${\cal R} \gg l_P$, the fundamental scale for gravity
$M_*$ can be significantly lower than the 4-dimensional one. By inverting the
above relation and using the definition $G_D=1/M_*^{n+2}$ for the fundamental
gravitational constant, we find
\begin{equation}
G_4 \,\,{\cal R}^{\,n} \simeq G_D\,.
\label{kanNewton}
\end{equation}
This means that, while, for $r \gg {\cal R}$, the Newtonian potential between
two masses $m_1$ and $m_2$ is given by the well-known 4-dimensional formula
\begin{equation}
V(r)=G_4\,\frac{m_1 m_2}{r}\,,
\end{equation}
for $r \ll {\cal R}$, the corresponding potential is now not only a 
higher-dimensional one but a much stronger one for the same masses $m_1$
and $m_2$, and is written as
\begin{equation}
V(r)=G_D\,\frac{m_1 m_2}{r^{n+1}}\,.
\end{equation}

In the case of the scenario with Warped Extra Dimensions, shown in Fig. \ref{kanbranes}(b),
a 4-dimensional brane is embedded in the higher-dimensional spacetime which now is
five-dimensional. The extra spacelike dimension is generically non-compact but it may
be compactified at will if a second brane is introduced in the model. The {\it visible}
brane, where all SM fields live, is placed at a finite distance $y=L$ from the  
{\it hidden} brane located at $y=0$. If all fundamental scales at the hidden brane
are of the order of $M_*$, then it may be shown that the electroweak symmetry
breaking in the visible brane takes place at a scale \cite{kanRS}
\begin{equation}
M_{EW}=e^{-kL}\,M_*\,,
\end{equation}
where $k$ is the curvature scale associated with the negative cosmological
constant that fills the 5-dimensional spacetime of the model. The effective
Planck scale $M_P$ is now related to the fundamental one $M_*$ through the equation
\begin{equation}
M_P^2 = \frac{M_*^3}{k}\,(1-e^{-2 k L})\,.
\end{equation}

In both scenaria, a low-scale gravitational theory can be realized in the context
of the higher-dimensional model. As we will see, this will have important 
consequences for the creation and evaporation process of black holes in these
theories. In these lectures, we will concentrate on the scenario with Large Extra
Dimensions, however, many of the arguments and results that will be presented hold
for the scenario with Warped Extra Dimensions, too, under the assumption that the
AdS radius $1/k$ is much larger than the horizon radius $r_H$ of the corresponding
black holes.

\begin{center}
\begin{table}[t]  
\caption{Current limits on the fundamental energy scale}
\begin{tabular}{lrr} \hline\noalign{\smallskip}
\hspace*{0.2cm}Type of Experiment/Analysis &  \hspace*{0.7cm} $M_* \ge$ \hspace*{0.5cm} 
& \hspace*{0.7cm} $M_* \ge$ \hspace*{0.5cm}\\ 
\noalign{\smallskip}\svhline\noalign{\smallskip}
{\begin{tabular}{l}Collider limits on the production \\
of real or virtual KK gravitons \cite{kanDelphi}-\cite{kanCDF}\end{tabular}}
& \hspace*{1.0cm} 1.45 TeV ($n=2$) & \hspace*{1.0cm} 0.6 TeV ($n=6$)\\[3mm]  
Torsion-balance Experiments\cite{kanHoyle, kanKapner}  & 3.2 TeV ($n=2$) &  
(${\cal R} \leq 50\,\mu$m)\\[1mm]
Overclosure of the Universe\cite{kanHall}  & 8 TeV ($n=2$) & \\[1mm]
Supernovae cooling rate \cite{kanCullen}-\cite{kanReddy2}  & 30 TeV ($n=2$)
&  2.5 TeV ($n=3$)\\[1mm]
Non-thermal production of KK modes \cite{kanPospelov}  & 35 TeV ($n=2$) & 
3 TeV ($n=6$)\\[1mm]
Diffuse gamma-ray background \cite{kanHall, kanHR1, kanHann}  & 110 TeV ($n=2$) & 
5 TeV ($n=3$)\\[1mm]
Thermal production of KK modes \cite{kanHann} & 167 TeV ($n=2$) &  1.5 TeV ($n=5$)\\[1mm]
Neutron star core halo \cite{kanHR2}  & 500 TeV ($n=2$) & 30 TeV ($n=3$) \\[1mm]
Time delay in photons from GRB's \cite{kanGSS} & 620 TeV ($n=1$) & \\[1mm] 
Neutron star surface temperature \cite{kanHR2} & 700 TeV ($n=2$) &  
0.2 TeV ($n=6$)\\[1mm]
BH absence in neutrino cosmic rays \cite{kanFeng} & & 1-1.4 TeV ($n \geq 5$)\\
\noalign{\smallskip}\hline\noalign{\smallskip} \end{tabular} 
\label{kanlimits}
\end{table}
\end{center}

\subsection{Creation of Black Holes}
\label{kancreation}

A summary of the most important -- experimental, astrophysical and cosmological --
limits on the fundamental energy scale $M_*$ is presented in Table \ref{kanlimits}.
From its entries one may see that, in general, the constraints become more relaxed
as the number of additional spacelike dimensions increases. The most optimistic
case is the one where the higher-dimensional Planck scale $M_*$ is very close
to the TeV scale -- this case is still viable, however, one needs to introduce
at least 3 additional spacelike dimensions. In this version of the model the
hierarchy between the gravitational and the electroweak scale almost disappears. 
What is more important, the scale of quantum gravity, where gravitational and
SM interactions become of the same magnitude, approaches the energy scale where
present-day and future experiment operate. As a result, if $M_*$ is of the
order of a few TeV, then collider experiments with $E>M_*$ can probe the 
strong gravity regime and may witness the creation of heavy, extended objects! 

The following question therefore arises naturally: can we then produce a
black hole in a collider experiment on our brane?
The idea was put forward in \cite{kanBF} very soon after the formulation of the
two aforementioned models with extra dimensions. In there, it was argued that 
during a high-energy scattering process with $E>M_*$ and impact parameter $b$
between the colliding particles, the following two cases should be expected:
{\bf (i)} if $b> r_H(E)$, elastic and inelastic processes will take place,
dominated by the exchange of gravitons, while {\bf (ii)} if $b<r_H(E)$, a
black hole will be formed according to the Thorne's Hoop Conjecture \footnote{
... which says that ``A black hole is formed when a mass M gets compacted into a
region whose circumference in every direction is \,${\cal C} \leq 2\pi r_H(E)$''.
A higher-dimensional version of this conjecture was developed in \cite{kanNewHoop}
where the ``circumference'' was substituted by the ``area'' ${\cal V}_{D-3}$
of the $(D-3)$-dimensional ``surface'' that now needs to be ${\cal V}_{D-3} 
\leq G_D\,M$.} \cite{kanThorne}
and the colliding particles will disappear for ever behind the event horizon.
In the above, $r_H(E)$ is the Schwarzschild radius that corresponds to the 
center-of-mass energy $E$ of the colliding particles.

Since gravity is higher-dimensional, every gravitational object, including the
produced black hole, will be generically higher-dimensional. We thus expect
the black hole to form on but also to extend off our brane. Under the assumption
that the produced black hole has a horizon radius $r_H$ much smaller than the
size of the extra dimensions ${\cal R}$ -- a case that can be indeed realized as we will
see in the next subsection, it may be assumed that it lives in a spacetime
with $(4+n)$ non-compact dimensions. The simplest such black hole is the 
spherically-symmetric, neutral, higher-dimensional one described by the
Schwarzschild-Tangherlini line-element \cite{kanTangherlini}
\begin{equation}
ds^2 = - \left[1-\left(\frac{r_H}{r}\right)^{n+1}\right]\,dt^2 +
\left[1-\left(\frac{r_H}{r}\right)^{n+1}\right]^{-1}\,dr^2 + r^2 d\Omega_{2+n}^2\,,
\label{kanST}
\end{equation}
where $d\Omega_{2+n}^2$ is the line-element of a $(2+n)$-dimensional unit sphere
\begin{equation}
d\Omega_{2+n}^2=d\theta^2_{n+1} + \sin^2\theta_{n+1} \,\biggl(d\theta_n^2 +
\sin^2\theta_n\,\Bigl(\,... + \sin^2\theta_2\,(d\theta_1^2 + \sin^2 \theta_1
\,d\varphi^2)\,...\,\Bigr)\biggr)\,.
\label{unit}
\end{equation}
By applying the Gauss law in $D=4+n$ dimensions, we find for the horizon radius
the result \cite{kanMP}
\begin{equation}
r_H = {1\over M_*} \left(M_{BH}\over M_*\right)^{1\over n+1}
\left(8 \Gamma(\frac{n+3}{2}) \over (n+2) \sqrt{\pi}^{(n+1)}\right)^{1/(n+1)}\,.
\label{kanhorizon}
\end{equation}
The above expression reveals the, by now, well-known result that, in an arbitrary
number of dimensions, the horizon radius of the black hole has a power-law 
dependence on its mass $M_{BH}$ -- the more familial linear dependence is restored
if one sets $n=0$. More importantly, it is the fundamental Planck scale $M_*$
that appears in the denominator instead of the 4-dimensional one $M_P$, a feature
that will play an important role on deciding whether black holes may be created
at high-energy particle collisions. 

Turning therefore to this question, the basic criterion for the creation of such
a black hole is \cite{kanMR} that the Compton wavelength $\lambda_C=4\pi/E$ of the
colliding particle of energy $E/2$ must lie within the corresponding Schwarzschild
radius $r_H(E)$. By using the expression for the horizon radius (\ref{kanhorizon}),
the above is written as
\begin{equation}
\frac{4\pi}{E} < {1\over M_*} \left(E\over
M_*\right)^{1\over n+1} \left(8 \Gamma(\frac{n+3}{2}) \over (n+2)
\sqrt{\pi}^{(n+1)}\right)^{1/(n+1)}\,.
\label{kancriterion}
\end{equation}
This inequality can be solved to give the ratio $x_{min}=E/M_*$,  necessary
for the creation of the black hole. The results for $x_{min}$ for various values
of the number of extra dimensions $n$ are given in Table \ref{kanxmin}.
From these, we conclude that the center-of-mass energy of the collision must
be approximately one order of magnitude larger than the fundamental Planck
scale $M_*$. Note that, if the factor $4\pi$ is left out, as it was often done
in earlier back-on-the-envelope calculations, the constraint on $E$ comes out
to be much more relaxed, i.e. $E \geq M_*$. As the maximum center-of-mass energy
that can be achieved at the Large Hadron Collider at CERN is 14 TeV, it seems
that a window of approximately 5 TeV remains at our disposal to witness a
strong gravity effect such as the creation of a black hole.

\begin{table}[t]
\caption{The values of the ratio $x_{min}=E/M_*$, necessary for the creation of 
a black hole, as a function of $n$.}
\begin{tabular}{cccccc}  \hline\noalign{\smallskip}
$n=2$\, & \,$n=3$ \,& \,$n=4$\, & \,$n=5$ \,& \,$n=6$\, & \,$n=7$ \\
\noalign{\smallskip}\svhline\noalign{\smallskip}
\hspace*{0.05cm} $x_{min}=8.0$ \hspace*{0.2cm} & \hspace*{0.2cm} $x_{min}=9.5$ \hspace*{0.2cm}
& \hspace*{0.2cm} $x_{min}=10.4$ \hspace*{0.2cm} & \hspace*{0.2cm} $x_{min}=10.9$
\hspace*{0.2cm} & \hspace*{0.2cm} $x_{min}=11.1$ \hspace*{0.2cm} & 
\hspace*{0.2cm} $x_{min}=11.2$ \hspace*{0.05cm} \\
 \noalign{\smallskip}\hline\noalign{\smallskip} \end{tabular} 
 \label{kanxmin}
\end{table}

Moving beyond the classical criterion (\ref{kancriterion}) that allows for the
formation of the black hole, two basic questions arise next:
{\bf (i)} what {\it part} of the available center-of-mass energy $E$ is absorbed
inside the black hole, and {\bf (ii)} how {\it likely} is the creation of a black
hole at the first place. In order to answer these questions, we need to study
the details of the high-energy particle collision in a strong gravitational
background. A theory of Quantum Gravity could provide the answers, however, 
such a theory -- in a complete, consistent form -- is still missing. Over the
years the fast-moving, colliding particles have been modelled by gravitational 
waves, shock waves, and strings in the context of different theories such as
General Relativity (with or
without Quantum mechanics) \cite{kankhan, kanszekeres, kandray, kanyurtsever,
kanDP, kanthooft, kanGS}, String Theory \cite{kangross, kanamati, kanGGM} and Topological
Field Theory \cite{kanverlinde}. The most widely established method is the
use of the concept of the Aichelburg-Sexl shock wave \cite{kanAS} that was developed more
than twenty years ago in the context of a 4-dimensional gravitational theory.
An  Aichelburg-Sexl shock wave follows from a Schwarzschild line-element
boosted along the $z$ axis, with a Lorentz factor $\gamma=1/\sqrt{1-\beta^2}$. In
the limit $\gamma \rightarrow \infty$, the boosted line-element becomes 
\cite{kanDP}
\begin{equation}
ds^2=-du dv + dx^2 + dy^2 +4\mu \ln(x^2+y^2) \,\delta(u)\,du^2\,,
\label{kanASwave}
\end{equation}
with $u=t-z$ and $v=t+z$, and $\mu$ the particle's energy. 
The above line-element is everywhere {flat} apart from the point $u=0$ where
a discontinuity arises. Therefore, it describes a shock wave located at this
point and moving along the $+z$ axis at the speed of light. 

\begin{figure}[t]
\sidecaption
\includegraphics[scale=0.60]{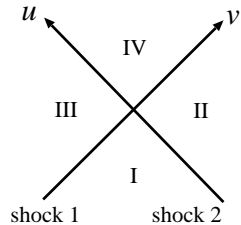}
\caption{Two Aichelburg-Sexl shock waves propagating at opposite directions.
The two shock waves collide at $u=v=0$ and, if a closed trapped surface or
an apparent horizon is formed, a black hole has been created.}
\label{kanASplot}
\end{figure}

We now assume that two Aichelburg-Sexl waves, with their centers at $u=0$
and $v=0$, are moving at opposite directions, one along the $+z$ axis and
the other along the $-z$. Then, the two shock waves will collide at 
$u=v=0$, as we may see at Fig. \ref{kanASplot}. The points at regions I,
II and III, that lie away from the moving trajectories and the collision
point, are flat, however, the region IV, which forms after the collision
is highly non-linear and curved\footnote{The task of finding the exact form
of spacetime in region IV involves strong, and thus non-linear, gravitational
calculations; until today no answer -- analytical or numerical -- has been
given to this question.}. If, at the union of the two shock waves, a 
{\it closed trapped surface} (i.e. a closed 2-dimensional spacelike surface
on which the outgoing orthogonal null geodesics have positive convergence
\cite{kanGeneral}) or an {\it apparent horizon} (that is, a closed trapped
surface with exactly zero convergence) is formed, then a black hole has
been created -- since, according to the Cosmic Censorship Hypothesis, 
the apparent horizon either coincides with or lies inside the event horizon
\cite{kanGeneral}.

The creation therefore of a black hole is nothing but a Boundary Value problem. 
In $D=4$ dimensions and for a head-on collision ($b=0$), this problem can be
solved analytically. This task was performed by Penrose \cite{kanPenrose},
more than 30 years ago, who found that an apparent horizon is indeed formed
with an area equal to $32 \pi \mu^2$. This can put a lower bound on the area of
the event horizon and thus on the black hole mass as follows:
\begin{equation}
A_H \equiv 4\pi r_H^2 \geq 32 \pi \mu^2 \,{\Rightarrow}\, M_{BH}\equiv
\frac{r_H}{2} \geq \frac{1}{\sqrt{2}}\,(2\mu)\,.
\end{equation}
From the above one may conclude that at least 71\% of the initial energy $E=2\mu$
of the collision is trapped inside the black hole.
Alternatively, one may compute the amount of energy emitted in the form of
gravitational waves during the violent collision. This was done in \cite{kanDP}
where it was found that that amount was of the order of 16\%, which
raised the percentage of energy absorbed by the black hole to 84\% of the
initial energy $E$. In more recent years, numerical analyses, where
one \cite{kanCardoso1} or both \cite{kanCardoso3} of the colliding bodies
were assumed to be described by a black hole, have found that the percentage
of energy lost in the form of gravitational waves is in the area of 14\%,
whish is in very good agreement with the results of \cite{kanDP}.

In the brane-world scenario, the colliding particles need to enter the
higher-dimensional regime in order to create a black hole. In that case,
every closed trapped surface will be a $(D-2)$ surface instead of a 
2-dimensional. Nevertheless, the same procedure for investigating the creation
of a black hole can be followed in this case, too. For a head-on collision,
the corresponding boundary value problem can be again solved analytically
leading to \cite{kanEG}
\begin{equation}
M_{BH} \geq [{ 0.71}\,({\rm for}\,D=4) - { 0.58}\,({\rm for}\,D=11)]\,(2\mu)\,.
\end{equation}
Therefore, as the dimensionality $D$ of spacetime increases, smaller and smaller
black holes will be created. On the other hand, if the collision is not head-on,
i.e. $b \neq 0$, numerical means have to be used to find the solution of the problem.
This was done in \cite{kanEG, kanYN} leading respectively to the results 
\begin{itemize}
\item $D=4\,$: \qquad \,\,\,$b \leq b_{max} \simeq 0.8\,r_H\,,$ \\
\item $D=4+n\,$: \,\, $b \leq b_{max} \simeq  3\,\,2^{-(n+2)/(n+1)}\,r_H\,.$ 
\end{itemize}
Thus, for a non-head-on collision, a black hole will be created if the impact
parameter is smaller than a fraction of the event horizon radius. This fraction
is 0.8 in $D=4$ and increases, reaching asymptotically unity, as $D$ becomes
larger. 

The impact parameter can offer us a measure of how likely the creation of
a black hole is. For a high-energy collision with a non-zero impact parameter
$b$, the production cross-section is found by using the geometric limit
\begin{equation}
\sigma_{\rm production} \simeq \pi b^2\,.
\label{kanproductioncross}
\end{equation}
According to the above, the cross-section for the production of black holes
from two fast-moving particles is assumed to be given by the classical
formula for the ``target'' area defined by the impact parameter. One
might intuitively think that the formation of a black hole at such high
energies would be governed by quantum, rather than classical, effects
-- in \cite{kanVoloshin}, the argument that the production cross-section
would be suppressed by an exponential factor involving the Euclidean
action of the system was put forward. However, in subsequent studies
\cite{kanDimEmp, kanKolVen, kanSolodukhin, kanRizzo, kanHsu} it was
shown that the creation of the black hole was a classically allowed
process and not a quantum phenomenon; the main contribution to the
production cross-section is therefore given by the classical expression
(\ref{kanproductioncross}) with the quantum corrections being indeed
small as in \cite{kanVoloshin}.

Early studies used the approximate expression $\sigma_{\rm production} \simeq
\pi r_H^2$, but today a more precise expression is needed. As we just saw, even
in the $D=4$ case, a black hole will not be created unless the impact parameter
is smaller than $0.8\,r_H$, a result that leads to the more accurate estimate
for the production cross-section: $\sigma_{\rm production} \simeq 0.64\,(\pi r_H^2)$.
Nevertheless, we are not done: novel estimates for the production cross-section
have emerged from the study of Ref. \cite{kanYR}, where the search for the creation
of a closed trapped surface was extended in the regime
($u=0$, $v>0$) and ($u>0$, $v=0$), i.e. in the `future' of the collision point.
This extension gave a boost to the production cross-section, since in cases
where it was previously concluded that no event horizon had been formed at the
collision point, now such a surface was found when the extended regime was used
instead. Therefore, the state-of-the-art values of the black hole production
cross-section are the ones given in Table \ref{kanYRcross} \cite{kanYR}. For
example, the $D=4$ value of 0.64, in units of $\pi r_H^2$, has increased to
0.71, with similar or larger enhancements taking place for the other values
of $D$, too\footnote{We should note here that the use of the generalized
uncertainty principle has shown to lead to an increase in the minimum amount
of energy needed for the creation of a black hole \cite{kanDas}. Similarly,
the production cross-section comes out to be suppressed if the charge of the
colliding particles exceeds a certain value \cite{kanYoshMann}, while the
angular momentum of the black hole enhances $\sigma_{\rm production}$
\cite{kanIOP1}. Finally, if one assumes the existence of a non-Gaussian
point in General Relativity and thus a running gravitational coupling,
the black-hole production cross-section is greatly suppressed in part
of the parameter space \cite{kanKoch}.}. 

\begin{table}[t]
\caption{Black-Hole production cross-section as a function of the
dimensionality of spacetime \cite{kanYR}}
\begin{tabular}{ccccccccc} \hline\noalign{\smallskip}
$D$ \hspace*{0.4cm}& \hspace*{0.4cm} 4 \hspace*{0.4cm} & \hspace*{0.4cm} 5 
\hspace*{0.4cm} & \hspace*{0.4cm} 6 \hspace*{0.4cm} & \hspace*{0.4cm} 7 \hspace*{0.4cm}
& \hspace*{0.4cm} 8 \hspace*{0.4cm} & \hspace*{0.4cm} 9 \hspace*{0.4cm} &
\hspace*{0.4cm} 10 \hspace*{0.4cm} & \hspace*{0.4cm} 11 \hspace*{0.1cm} \\
\noalign{\smallskip}\svhline\noalign{\smallskip}
$\sigma_{\rm production}/(\pi r_H^2)$ & 0.71 & 1.54 & 2.15 & 2.52 & 2.77 & 2.95
& 3.09 & 3.20 \\
\noalign{\smallskip}\hline\noalign{\smallskip} \end{tabular} 
\label{kanYRcross}
\end{table}

Focusing, for a moment, on the geometrical instead of the numerical factor in the
expression of the production cross-section, we may write
\begin{equation}
\sigma_{\rm production} \propto \pi r_H^2 \sim 
\frac{1}{M_*^2}\,\left(\frac{E}{M_*}\right)^{2/(n+1)}\,.
\end{equation}
The above expression gives the dependence of the production cross-section
on the center-of-mass energy of the collision and reveals the enhancement
of it with $E$, a dependence that is not seen in any other SM or Beyond
the Standard Model process\footnote{For a black hole produced in a 5-dimensional
Anti de Sitter spacetime, the above result for the production cross-section
holds if we assume that $r_H \ll 1/k$; on the other hand, if $r_H \ge 1/k$,
then, the corresponding expression for the production cross-section is
$\sigma_{\rm production} \propto \ln^2 E$ \cite{kanGiddings}.}. The above expression
is valid for the production of a black hole out of two elementary, non-composite
particles (i.e. partons). The final result for the production cross-section
out of two accelerated composite particles, such as protons,
follows by properly summing over all pairs of partons that carry enough energy
to produce a black hole. This is finally given by \cite{kanGT, kanDL} 
\begin{equation}
\sigma_{\rm production}^{\,pp \rightarrow BH}=\sum_{ij} \int_{\tau_m}^1\,d\tau\,
\int_{\tau}^{1} \frac{dx}{x}\,f_i(x)\,f_j(\frac{\tau}{x})\,\sigma_{\rm production}^{ij
\rightarrow BH}\,,
\end{equation}
where $x$ is the parton-momentum fraction, $\tau=\sqrt{x_i\,x_j}$, and $f_i(x)$ are the
so-called parton distribution functions (PDF's) that determine the fraction of the
center-of-mass energy that is carried by the partons. 

Summing over all possible pairs of partons gives another considerable boost to
the production cross-section. One could naively think that by increasing without
limit the available center-of-mass energy $E$, one could create extremely
energetic pairs of partons each one of which would certainly create a 
black hole. However, the parton distribution functions $f_i(x)$ decrease rapidly
with the center-of-mass energy $E$, and with them the amount of energy that is
passed to (and retained by) the partons. As a result, the production cross-section
can not be indefinitely increased. Numerical calculations, that take into
account the compositeness of the accelerated particles and the behaviour of 
PDF's, have derived some indicative values for  $\sigma_{\rm production}$
\cite{kanGT, kanDL}.
For example, if we assume that $M_*=1$ TeV and $D=10$, then the production cross-section
for a black hole with $M_{BH} = 5$ TeV turns out to be $\sigma_{\rm production}
\sim 10^5$ fb, while for a black hole with $M_{BH} = 10$ TeV it is found that
$\sigma_{\rm production} \sim 10$ fb. For beyond the SM processes, the aforementioned
values are quite significant -- in the first case, the value of $\sigma_{\rm production}$
amounts to one black hole created per second! Whether LHC will indeed prove to be
a black-hole factory, it remains to be seen. 


\subsection{Black-Hole Properties}
\label{kanproperties}

We now turn to the properties of the higher-dimensional black holes that may
be produced during trans-planckian particle collisions \cite{kanADMR, kanreview}.
We will use as a prototype for our discussion the spherically-symmetric,
neutral black hole described by the Schwarzschild-Tangherlini line-element 
(\ref{kanST}). Let us start with the
value of the horizon radius -- how big (or, small) are actually these black
holes? The value of $r_H$ as a function of the mass of the black hole is
given in Eq. (\ref{kanhorizon}). In order to derive some realistic estimates,
we assume again $M_*=1$ TeV and $M_{BH}=5$ TeV, and calculate the value of
the horizon as a function of the number of extra dimensions $n$. These
values are presented in Table \ref{kanprop}. From these we may easily
conclude that, in the presence of extra dimensions, in order to create a
black hole we only need to access subnuclear distances. To have a measure
of comparison, let me note that, in $D=4$ with $M_*=M_P \simeq 10^{19}$ GeV,
the same objective could only be achieved if the two colliding particles
came within a distance of $10^{-35}$ fm!

\begin{table}[t]
\caption{Horizon radius and temperature of the Schwarzschild-Tangherlini black 
hole as a function of the number of extra dimensions, for $M_*=1$\,TeV and
$M_{BH}=5$\,TeV}
\begin{tabular}{cccccccc} \hline\noalign{\smallskip}
\hspace*{0.5cm} $n$ \hspace*{0.5cm}& \hspace*{0.5cm} 1 \hspace*{0.5cm} & \hspace*{0.5cm} 2 
\hspace*{0.5cm} & \hspace*{0.5cm} 3 \hspace*{0.5cm} & \hspace*{0.5cm} 4 \hspace*{0.4cm}
& \hspace*{0.5cm} 5 \hspace*{0.5cm} & \hspace*{0.5cm} 6 \hspace*{0.5cm} &
\hspace*{0.5cm} 7 \hspace*{0.5cm} \\
\noalign{\smallskip}\svhline\noalign{\smallskip}
$r_H$ ($10^{-4}$ fm) \hspace*{0.0cm} & 4.06 & 2.63 & 2.22 
& 2.07 & 2.00 & 1.99 & 1.99 \\[1mm] 
$T_{H}$ (GeV) & 77 & 179 & 282 & 379 & 470 
& 553 & 629 \\
\noalign{\smallskip}\hline\noalign{\smallskip}
\end{tabular}
\label{kanprop}
\end{table}

Drawing from our knowledge of their 4-dimensional analogues, we expect that
these miniature black holes will go through the following stages during their
lifetime \cite{kanGT}: {\bf (i)} The balding phase: the initially highly
asymmetric black hole will shed all quantum numbers and multipole moments
apart from its mass $M$, electromagnetic charge $Q$, and angular momentum $J$
-- during this phase, we expect some visible but mainly invisible energy emission.
{\bf (ii)} The spin-down phase: the black hole will start losing its angular
momentum via the emission of Hawking radiation through mainly visible channels.
{\bf (iii)} The Schwarzschild phase: after its angular momentum the black hole
will now start losing its mass through the emission again of Hawking radiation.
{\bf (iv)} The Planck phase: when  the black hole $M_{BH}$ approaches $M_*$, it becomes
a quantum object whose properties would follow only from a quantum theory of
gravity -- possible scenaria for this phase are the emission of a few energetic
quanta leading to the complete evaporation of the black hole, or the formation
of a stable ``quantum'' remnant. 

The emission of Hawking radiation \cite{kanHawking} is sourced by the non-vanishing
temperature of the black hole. This is defined in terms of the black hole's
surface gravity $k$ as follows
\begin{equation}
T_{H} = \frac{k}{2\pi}=
\frac{1}{4\pi}\,\frac{1}{\sqrt{|g_{tt}\,g_{rr}|}}\left(\frac{d|g_{tt}|}{dr}
\right)_{r=r_H}={(n+1) \over 4\pi\,r_H}\,.
\label{kantemp}
\end{equation}
By using again, as an indicative case, the values $M_*=1$\,TeV and $M_{BH}=5$\,TeV,
for the fundamental Planck scale and black-hole mass, as well as the values of
the horizon radius $r_H$ given in Table \ref{kanprop}, we may calculate the
temperature of the black hole in terms of the number of extra dimensions. These
are also given in Table \ref{kanprop}. We observe that a higher-dimensional black
hole, with mass in the range of values that would allow it to be produced at LHC, 
comes out to have in addition a temperature that would greatly facilitate its
detection at present and future experiments -- unlike the large
astrophysical black holes that are characterised by an extremely low temperature
and the majority of primordial black holes that have an extremely high temperature.
Finally, let us add that due to the emission of Hawking radiation, the lifetime
of a black hole is finite. In the case of a higher-dimensional black hole, this
quantity is given by \cite{kanADMR}
\begin{equation}
\tau_{(n+4)} \sim \frac{1}{M_*}\left(\frac{M_{BH}}{M_*}\right)^{\frac{(n+3)}{(n+1)}}
> \tau_{(4)}\,.
\end{equation}
For the same values of $M_*$ and $M_{BH}$, the typical lifetime of the black hole
comes out to be $\tau=(1.7-0.5) \times 10^{-26}$ sec for $n=1-7$. In other words, 
the produced black hole will evaporate instantly after its creation, and it
will do so right in front of our detectors.  

That is why we need to study in the greatest possible detail the spectrum of the
Hawking radiation emitted by the black hole as this will probably be the main
observable effect associated with this gravitational object. Although a purely
geometrical property, the temperature of a black hole leads to the emission of
thermal radiation similar to that of a black body. The Hawking radiation \cite{kanHawking}
is therefore a classical phenomenon but with a quantum origin, since classically
nothing is allowed to escape from within the black-hole horizon. The emission
of radiation from a black hole, 4-dimensional and high-dimensional alike, can
be realized through the creation of a virtual pair of particles just outside
the horizon; when the antiparticle happens to fall inside the black hole, the particle
can then propagate away from the black hole whose mass has decreased due to
the negative amount of energy it received. The radiation spectrum is therefore
a nearly black-body spectrum with energy emission rate given by an expression
of the form \cite{kanHawking}
\begin{equation}
\frac{\textstyle dE(\omega)}
{\textstyle dt} = \frac{\textstyle {  |{\cal A}(\omega)|^2} \,\,\omega}
{\textstyle \exp\left(\omega/T_{H}\right) \mp 1}\,\,
\frac{\textstyle d\omega}{\textstyle (2\pi)}\,.
\end{equation}
The quantity${|{\cal A}(\omega)|^2}$ appearing in the numerator is the Absorption 
Probability (or, {\it greybody factor}). Its presence is due to the fact that
a particle,  propagating in the $(4+n)$-dimensional black-hole background, needs
to escape the strong gravitational field, that the black hole creates, to reach
the asymptotic observer. In order to see this, we may write the equation of motion
of an arbitrary field in the aforementioned background in the form of a 
Schr\"odinger-like equation
\begin{equation}
-\frac{d^2 \Psi}{dr_*^2} + V(r_*, n,l,\omega, s,...)\Psi = \omega^2\,\Psi\,,
\label{kanSchr}
\end{equation}
in terms of the so-called tortoise coordinate
$dr_*=\left[1-\left(\frac{r_H}{r}\right)^{(n+1)}\right]^{-1} dr$. The gravitational
barrier $V(r_*, n,l,\omega, s,...)$ will reflect some particles back to the black hole
while it will allow others to escape to infinity. The rate at which particles and
therefore energy is ``arriving'' at the location of the asymptotic observer will
thus be proportional to the transmission (or, absorption, as we will shortly see)
probability and thus different from the one for the usual, flat-space blackbody
radiation. However, the extra difficulty that the greybody factor introduces in
the calculation of the radiation spectrum is compensated by the following fact:
the barrier, and
consequently the absorption probability, depends on a number of parameters that
describe both particle properties (spin $s$, energy $\omega$, angular momentum
numbers $l$,$m$, ...) and spacetime properties (number of extra dimensions $n$,
angular momentum of black hole $a$, cosmological constant $\Lambda$, ...). As a
result, the Hawking radiation spectrum, when computed, is bound to be a vital
source of information on the emitted particles and gravitational background. 

But how does the radiation spectrum follow? For this, we need to do a Quantum
Field Theory analysis in curved spacetime. The first step is to define a basis
for our fields: in the 4-dimensional case, we write \cite{kanUnruh, kanOW} 
\begin{equation}
u_{\omega l m} =\frac{N}{r}\,e^{-i\omega t}\,e^{i m \varphi}\,
S_{\omega l m}(\theta)\,R_{\omega l m}(r)\,,
\label{kanbasis}
\end{equation}
where $N$ is a normalization constant, ($l$, $m$) the angular momentum numbers with
$|m| \leq l$, and $S_{\omega l m}(\theta)$ the spherical harmonics. We also need
to define the vacuum state of the theory. The one that describes perfectly the
Hawking radiation emission process is the {\it past Unruh vacuum} $|U^- \rangle$:
this state has no incoming radiation from past null infinity ${\cal J}^-$ (i.e.
far away from the BH at some asymptotic initial time) but modes can ``come out''
of the black hole. 

The gravitational potential $V$ that appears in the equation of motion of the
field propagating in the black-hole background has the form of a barrier: it
is localised, and vanishes at both the horizon and infinity. At these two asymptotic
regimes, Eq. (\ref{kanSchr}) can then be easily solved, and the radial part
of the field assumes the forms  
\begin{equation} 
R_{\omega l m}^{up}(r) \sim \left\{\begin{tabular}{ll} $e^{i\omega r_*} + 
A^{up}\,e^{-i\omega r_*}$\,, \,\, & $r \rightarrow r_H$ \\[2mm] $B^{up}\,e^{i\omega r_*}$\,,
& $r \rightarrow \infty$\end{tabular}\right.\,.
\label{kanasymptotic1}
\end{equation}
The solution is, as expected, a superposition of free plane-waves, where the
constants $A^{up}$ and $B^{up}$ can be viewed as the Reflection and Transmission
coefficients. 

The fluxes of particles $N$ and energy $E$ emitted by the black hole and measured by
an observer at infinity are given by the vacuum expectation values of the radial
component of the conserved current $J^\mu$ and the $(tr)$-component of the
energy-momentum tensor $T_{\mu\nu}$, respectively, evaluated at infinity 
\cite{kanUnruh}
\begin{equation}
\frac{d^2\,\,\,}{r^2\,dt\,d\Omega}{\Biggl(\begin{tabular}{c} $N$\\ $E$
\end{tabular}\Biggr)}= \langle U^-|{\Biggl(\begin{tabular}{c} $J^r$\\ $T^{tr}$
\end{tabular}\Biggr)}|U^- \rangle_\infty\,. 
\label{kanfluxes}
\end{equation}
Using the asymptotic form (\ref{kanasymptotic1}) for the radial part of the
field at infinity, and after some algebra, we find
\begin{equation}
\frac{d^2\,\,\,}{\,dt\,d\omega}{ \Biggl(\begin{tabular}{c} $N$\\$E$
\end{tabular}\Biggr)}= \frac{1}{2\pi}\,\sum_{l}\,\,\frac{N_l\,|B^{up}|^2}{\exp(\omega/T_H) \mp 1}
\,{\Biggl(\begin{tabular}{c} $1$\\$\omega$
\end{tabular}\Biggr)}\,,
\label{kanflux}
\end{equation}
where $N_l=2l+1$ is the multiplicity of states that have the same value of
the angular momentum number $l$, and the $\pm1$ factor is a statistics factor
for fermions and bosons, respectively. 

We note that, in the numerator of the above expression, it is the transmission
probability $|B^{up}|^2$ that appears, as expected. However, one may define an
alternative, but equivalent, basis, namely
\begin{equation}
R_{\omega l m}^{in}(r) \sim \left\{\begin{tabular}{ll} $B^{in} e^{-i\omega r_*}$\,, \,\,
& $r \rightarrow r_H$ \\[2mm] $e^{-i\omega r_*} + A^{in}\,e^{i\omega r_*}$\,,
& $r \rightarrow \infty$\end{tabular}\right.\,.
\end{equation}
This basis describes modes that originate not from the black hole but from the
past null infinity. Now, $A^{in}$ and $B^{in}$ can be viewed as the 
Reflection and Absorption coefficients, respectively. As both sets of solutions
satisfy the same radial equation, one may easily show that the following
relations hold
\begin{equation}
1-|A^{in}|^2=|B^{in}|^2 \equiv |B^{up}|^2=1-|A^{up}|^2\,.
\end{equation}
From the above, we may easily conclude that the transmission probability
$|B^{up}|^2$ for the ``up'' modes originating from inside the black hole
is equal to the absorption probability $|B^{in}|^2$ for the ``in'' modes
originating from past null infinity -- we denote these two quantities
collectively as $|{\cal A}(\omega)|^2$, and write
\begin{equation}
\frac{d^2\,\,\,}{\,dt\,d\omega}{ \Biggl(\begin{tabular}{c} $N$\\$E$
\end{tabular}\Biggr)}= \frac{1}{2\pi}\,\sum_{l}\,\,\frac{N_l\,|{\cal A}(\omega)|^2}
{\exp(\omega/T_H) \mp 1}\,{\Biggl(\begin{tabular}{c} $1$\\$\omega$
\end{tabular}\Biggr)}\,.
\label{kanflux1}
\end{equation}

In the case of a rotating (Kerr) black hole, we may compute three rates:
the emission rates of particles $N$ and energy $E$ and the rate of loss of the angular
momentum $J$ of the black hole. These are given by the expressions
\begin{equation}
\frac{d^2\,\,\,}{r^2\,dt\,d\Omega}{\Biggl(\begin{tabular}{c} $N$\\$E$\\$J$
\end{tabular}\Biggr)}= \langle U^-|{\Biggl(\begin{tabular}{c} $J^r$\\$T^{tr}$\\$T^r_\varphi$
\end{tabular}\Biggr)}|U^- \rangle_\infty\,.
\end{equation}
The asymptotic solutions for the radial part of the field for either the ``up''
modes or the ``in'' modes propagating in a Kerr black-hole background are now
given by 
\begin{equation}
R_{\omega l m}^{up}(r) \sim \left\{\begin{tabular}{ll} $e^{i{  \tilde\omega} r_*} + 
A^{up}\,e^{-i{  \tilde\omega} r_*}$\,, \,\, & ${  r \rightarrow r_H}$ \\[2mm]
$B^{up}\,e^{i\omega r_*}$\,, & $r \rightarrow \infty$\end{tabular}\right.
\label{kanup}
\end{equation}
and
\begin{equation} 
R_{\omega l m}^{in}(r) \sim \left\{\begin{tabular}{ll} $B^{in} e^{-i{  \tilde \omega} r_*}$\,, \,\,
& ${  r \rightarrow r_H}$ \\[2mm] $e^{-i\omega r_*} + A^{in}\,e^{i\omega r_*}$\,,
& $r \rightarrow \infty$\end{tabular}\right.\,.
\label{kanin}
\end{equation}
In the above, the parameter $\tilde \omega$ is defined as
\begin{equation}
\tilde\omega  \equiv \omega - m\,\Omega_H = \omega -m \frac{a}{r_H^2+a^2}\,,
\end{equation}
where $\Omega_H$ is the angular velocity of the rotating black hole, and $a$ the
angular momentum parameter to be defined later. By using
as a basis the ``up'' modes, that, as we saw, describe more accurately the
Hawking radiation emission process, we find the expressions
\begin{equation}
\frac{d^2\,\,\,}{\,dt\,d\omega}{\Biggl(\begin{tabular}{c} $N$\\$E$\\$J$
\end{tabular}\Biggr)}= \frac{1}{2\pi}\sum_{l,m}\,\frac{\omega}{\tilde\omega}\,
\frac{|B^{up}|^2}{\exp(\tilde\omega/T_H)\mp 1}
\,{\Biggl(\begin{tabular}{c} $1$\\$\omega$\\$m$
\end{tabular}\Biggr)}\,.
\end{equation}
As in the non-rotating case, we also find that the following relations hold between
the coefficients of the asymptotic solutions (\ref{kanup})-(\ref{kanin}) for
the two sets of modes
\begin{equation}
\frac{\omega}{\tilde \omega}\,|B^{up}|^2 =1-|A^{up}|^2 { \equiv} 1-|A^{in}|^2=
\frac{\tilde\omega}{\omega}\,|B^{in}|^2\,,
\label{kan-relations}
\end{equation}
leading to the final, { simpler} formula for the three rates
\begin{equation}
\frac{d^2\,\,\,}{\,dt\,d\omega}{\Biggl(\begin{tabular}{c} $N$\\$E$\\$J$
\end{tabular}\Biggr)}= \frac{1}{2\pi}\sum_{l,m}\,
\frac{{  |{\cal A}(\omega)|^2}}{\exp(\tilde\omega/T_H)\mp 1}
\,{\Biggl(\begin{tabular}{c} $1$\\$\omega$\\$m$
\end{tabular}\Biggr)}\,,
\label{kan-ratesrot}
\end{equation}
where now ${  |{\cal A}(\omega)|^2} { \equiv}=1-|A^{up}|^2 { \equiv} 1-|A^{in}|^2$. 
Let us also note that if $\tilde \omega =\omega -m\,\Omega_H<0$, then from Eq.
(\ref{kan-relations}) the reflection probabilities $|A^{up}|^2$ and $|A^{in}|^2$
can be larger than unity - this happens only for modes with $m>0$ and signals
the effect of {\it superradiance} \cite{kanZeldovich}, where the incident wave
``steals'' energy from the rotating black hole and escapes with an amplitude
larger than the original one.

Let us now introduce a number of additional, spacelike dimensions in our theory.
Surprisingly, not much changes in the functional form of the above formulae. The
emission rates for a higher-dimensional, rotating black hole will still be given
by expressions of the form \cite{kanKMR1, kanKMR2, kanIOP1, kanFS, kanHK2, kanDHKW, 
kanIOP2, kanCKW, kanIOP3, kanCDKW}
\begin{equation}
\frac{d^2\,\,\,}{\,dt\,d\omega}{ \Biggl(\begin{tabular}{c} $N$\\$E$\\$J$
\end{tabular}\Biggr)}= \frac{1}{2\pi}\sum_{l,m,j...}\,
\frac{{  |{\cal A}(\omega)|^2}}{\exp(\tilde\omega/T_H)\mp 1}
\,{\Biggl(\begin{tabular}{c} $1$\\$\omega$\\$m$
\end{tabular}\Biggr)}\,.
\end{equation}
Where does the difference from the 4-dimensional case lie? To start with, the
temperature of the black hole will acquire an $n$-dependence. In addition, the
equation of motion of a given field is going to depend on the specific background;
therefore, the greybody factor $|{\cal A}(\omega)|^2$, that follows by solving
the corresponding equation of motion is going to change, too. Also, the symmetry
and structure of the higher-dimensional spacetime may introduce additional
quantum numbers and/or change the multiplicities of states that carry the same
sets of quantum numbers. 

Another important factor is whether we are considering emission of particles 
on the brane or in the bulk. Unlike a purely 4-dimensional black hole, a 
higher-dimensional one can emit particles either in the ``brane channel''
or in the ``bulk channel''. The species of particles that can be emitted in 
the bulk are particles that are allowed by the model to propagate in the
higher-dimensional spacetime, namely gravitons but also scalar fields that
carry no quantum numbers under the SM gauge group. These bulk modes ``see''
the full $(4+n)$-dimensional gravitational background and they are invisible to us,
therefore, any energy emitted in the bulk will be interpreted as a missing
energy signal for a brane observer. On the other hand, the black hole can
emit a variety of particles in the ``brane channel'', namely fermions, gauge
bosons and Higgs-like scalars. These brane-localised modes ``see'' only the
projected-on-the-brane 4-dimensional gravitational background and they are
directly visible to a brane observer; as a result, they are the most
interesting emission channel to study from the phenomenological point of view. 

%
%

\section{Second Lecture: Hawking Radiation Spectra and Observable Signatures}
\label{kanlecture2}

Having discussed the properties of the miniature black holes that may be
created during a high-energy particle collision in the context of a low-scale
higher-dimensional gravitational theory, we now proceed to discuss in more
detail the spectra of the Hawking radiation emitted by these black holes
and the information on particle and spacetime properties that we may deduce
from them.

\subsection{ The Schwarzschild Phase on the Brane}
\label{kanSchwarzbrane}

As we mentioned in the previous section, a black hole emits Hawking radiation
during the two intermediate phases of its life, namely the spin-down and the
Schwarzschild phase. We will start from the latter one, which although follows
the spin-down phase, was the first one to be studied due to the simpler form of
the line-element that describes the gravitational background around it. This
is given by the Schwarzschild-Tangherlini solution (\ref{kanST}) and describes,
as we have seen, a spherically-symmetric, neutral black hole that has lost all
of its angular momentum. For the purpose of studying the emission of Hawking
radiation directly on the brane, we will be interested in the brane-localised
modes that ``see'' only the projected-on-the-brane background. In order to
derive the latter, we fix the values of all the additional $\theta_i$ coordinates,
with $i=2,...,n+1$, introduced to describe the additional spacelike dimensions,
to $\frac{\pi}{2}$. Then, the resulting brane background assumes the form
\begin{equation}
ds^2_4 = - \left[1-\left(\frac{r_H}{r}\right)^{n+1}\right] dt^2 +
\left[1-\left(\frac{r_H}{r}\right)^{n+1}\right]^{-1} dr^2 + 
r^2\,d\Omega_2^2\,.
\label{kanprojectedSchw}
\end{equation}
The above line-element describes a 4-dimensional black-hole background on
the brane which, although resembles a Schwarzschild background, it is distinctly
different as it carries a non-trivial $n$-dependence. The horizon radius is
still given by Eq. (\ref{kanhorizon}) and its temperature by Eq. (\ref{kantemp})
-- note, that both the horizon radius and black-hole temperature follow from geometrical
arguments involving only the $g_{tt}$ and $g_{rr}$ metric components, and these are
not affected by the projection of the $(4+n)$-dimensional line-element onto the
brane. 

However, the different form of the gravitational background is bound to change
the equation of motion of the relevant species of particles, and thus the value
of the greybody factor $|{\cal A}(\omega)|^2$. In order to study in a combined way the
behaviour of fields with spin $s=0,1/2$ and 1, a ``master'' equation of motion
with $s$ appearing as a parameter was derived in \cite{kanKMR1, kanKMR2, kanreview}. 
For this, we used a factorized ansatz for the wavefuction of the field of the
form 
\begin{equation}
\Psi_s=e^{-i\omega t}\,e^{im\varphi}\,\Delta^{-s}\,R_s(r)\,S_{s l}^m(\theta)\,,
\end{equation}
and employed the Newman-Penrose method \cite{kanNP, kanChandra}, that combines
multi-component fields with curved gravitational backgrounds. Then, two decoupled
equations, one for the radial function $R_s(r)$ and one for the spin-weighted
spherical harmonics \cite{kanGoldberg} $S_{s l}^m(\theta)$, were derived
having the form
\begin{equation}
\Delta^s \frac{\textstyle d}
{\textstyle dr}\left(\Delta^{1-s}\frac{\textstyle d R_s^{~}}
{\textstyle dr}\right)+ \left[\frac{\textstyle \omega^2 r^2}{\textstyle h}+2i\omega s r-
\frac{\textstyle is\omega r^2 h'}{\textstyle h}
-\lambda_{s l}\right]R_s (r)=0\,,
\end{equation}
and
\begin{equation}
\frac{1}{\sin\theta}\,\frac{\textstyle d}
{\textstyle d\theta}\left(\sin\theta\frac{\textstyle d S^m_{s l}}
{\textstyle d\theta}\right)+ \left[-\frac{\textstyle 2ms\cot\theta}{\textstyle \sin\theta}
-\frac{m^2}{\sin^2\theta}+s-s^2\cot^2\theta+\lambda_{s l}\right]S^m_{s l}(\theta)=0\,,
\end{equation}
respectively. In the above, we have defined the function $\Delta\equiv r^2\,h
\equiv r^2\left[1- \left(\frac{r_H}{r}\right)^{n+1}\right]$, while
$\lambda_{s l}=l(l+1)-s(s-1)$ is the eigenvalue of the spin-weighted
spherical harmonics. The above equations resemble the ones derived by Teukolsky
\cite{kanTeukolsky} in the background of a purely 4-dimensional black-hole
background and differ only in the expressions of the functions $h(r)$ and $\Delta(r)$. 

The {radial} equation, from where the value of the greybody factor will follow,
may be solved either analytically or numerically. If the analytic approach is
chosen \cite{kanKMR1, kanKMR2}, an approximation method must be followed
according to which: {\bf (i)} we solve the equation of motion in the Near-Horizon
(NR) regime ($r \simeq r_H$) where it takes the form of a hypergeometric equation,
{\bf (ii)} then, we solve the equation of motion in the Far-Field (FF) regime 
($r \gg r_H$) where it takes the form of a confluent hypergeometric equation, and
{\bf (iii)} finally, we match the two asymptotic solutions in an intermediate zone
to guarantee the existence of a smooth solution over the whole radial regime.
Once the solution for the radial function $R_s(r)$ is found, we compute the 
Absorption Probability (we use the ``in'' modes as a basis) through the formula
\begin{equation}
|{\cal A}(\omega)|^2 \equiv 1-|{\cal R}(\omega)|^2 \equiv \frac{\textstyle 
\hspace*{2mm}{\cal F}_{\rm horizon_{~}}} 
{\textstyle {\cal F}^{~}_{\rm infinity}}\,,
\end{equation}
where ${\cal R}(\omega)$ is the Reflection coefficient and ${\cal F}$ the flux
of energy towards the black hole.

\begin{figure}[t]
\hbox{\scalebox{0.55} {\rotatebox{0}
{\includegraphics[width=10cm, height=7cm]{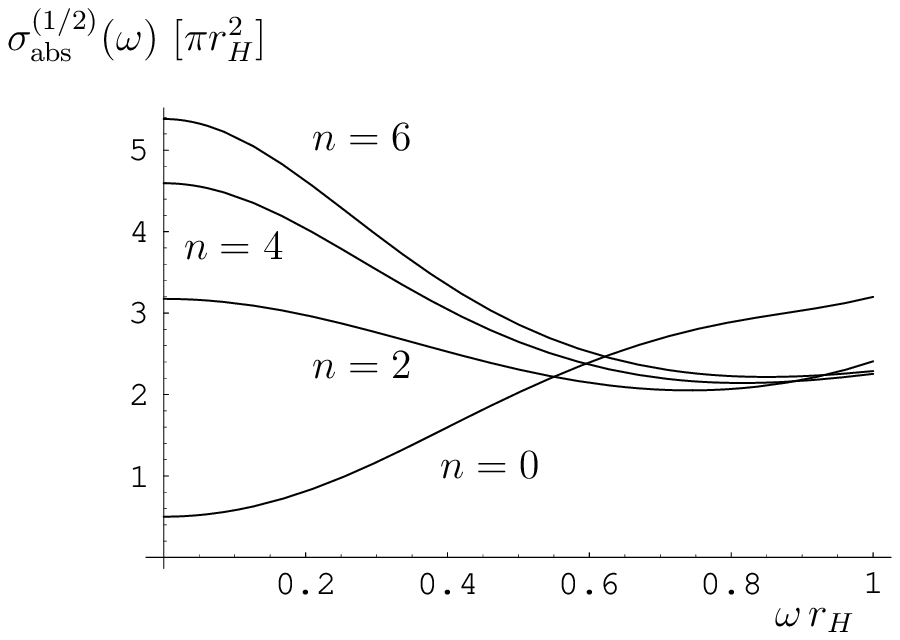}}}
\hspace*{4cm}}

\vspace*{-4.2cm}
\hspace*{5.5cm}\hbox{
\hspace*{0.4cm}\scalebox{0.55}
{\rotatebox{-90}{\includegraphics[width=7cm, height=10cm]{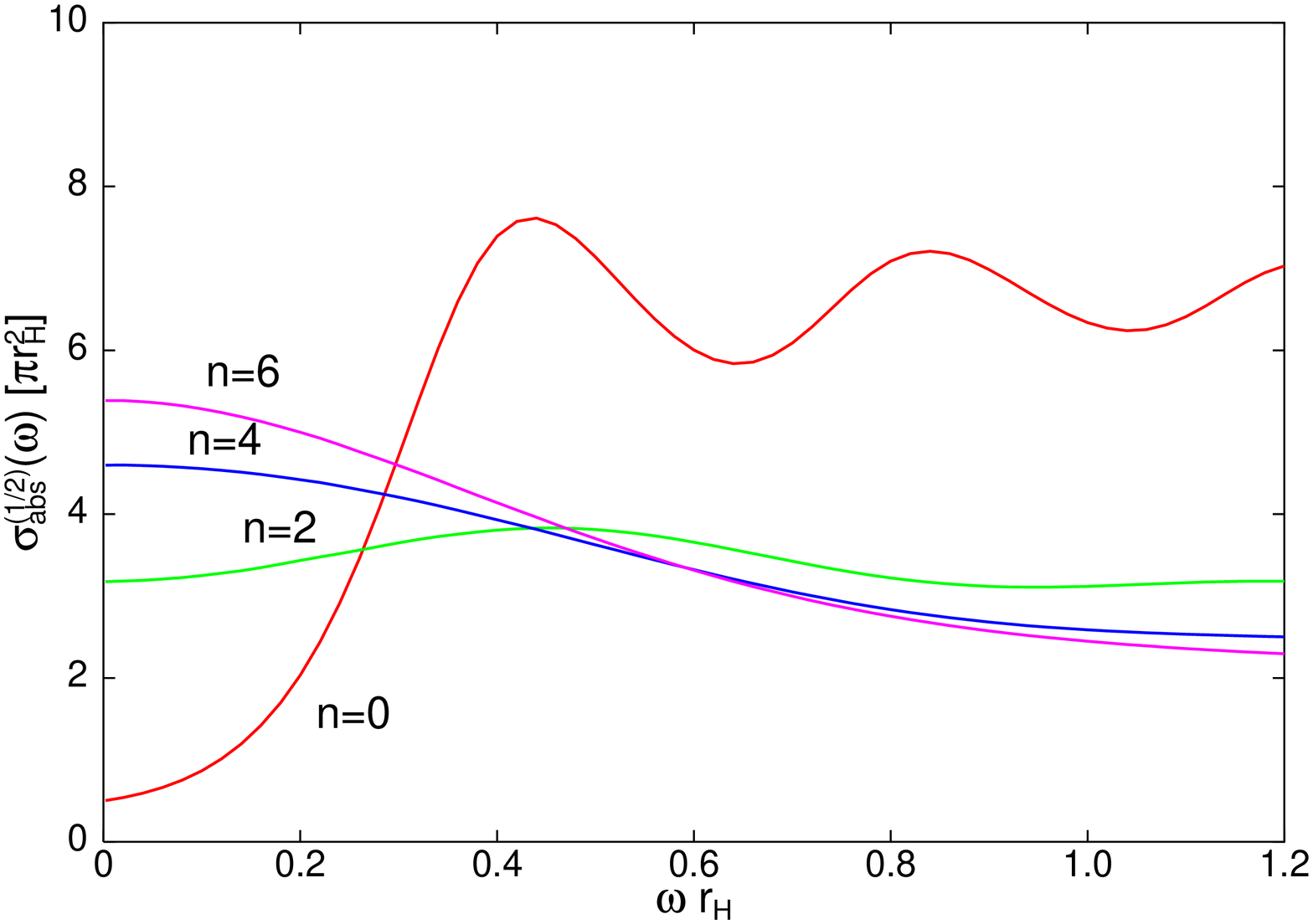}}}}
\caption{Absorption cross-section for brane-localised fermions evaluated analytically
(left plot), and numerically (right plot).}
\label{kansabsfer}
\end{figure}

\begin{figure}[t]  
\hspace*{0.0cm}\hbox{
\scalebox{0.55}{\rotatebox{-90} 
{\includegraphics[width=7cm, height=10cm]{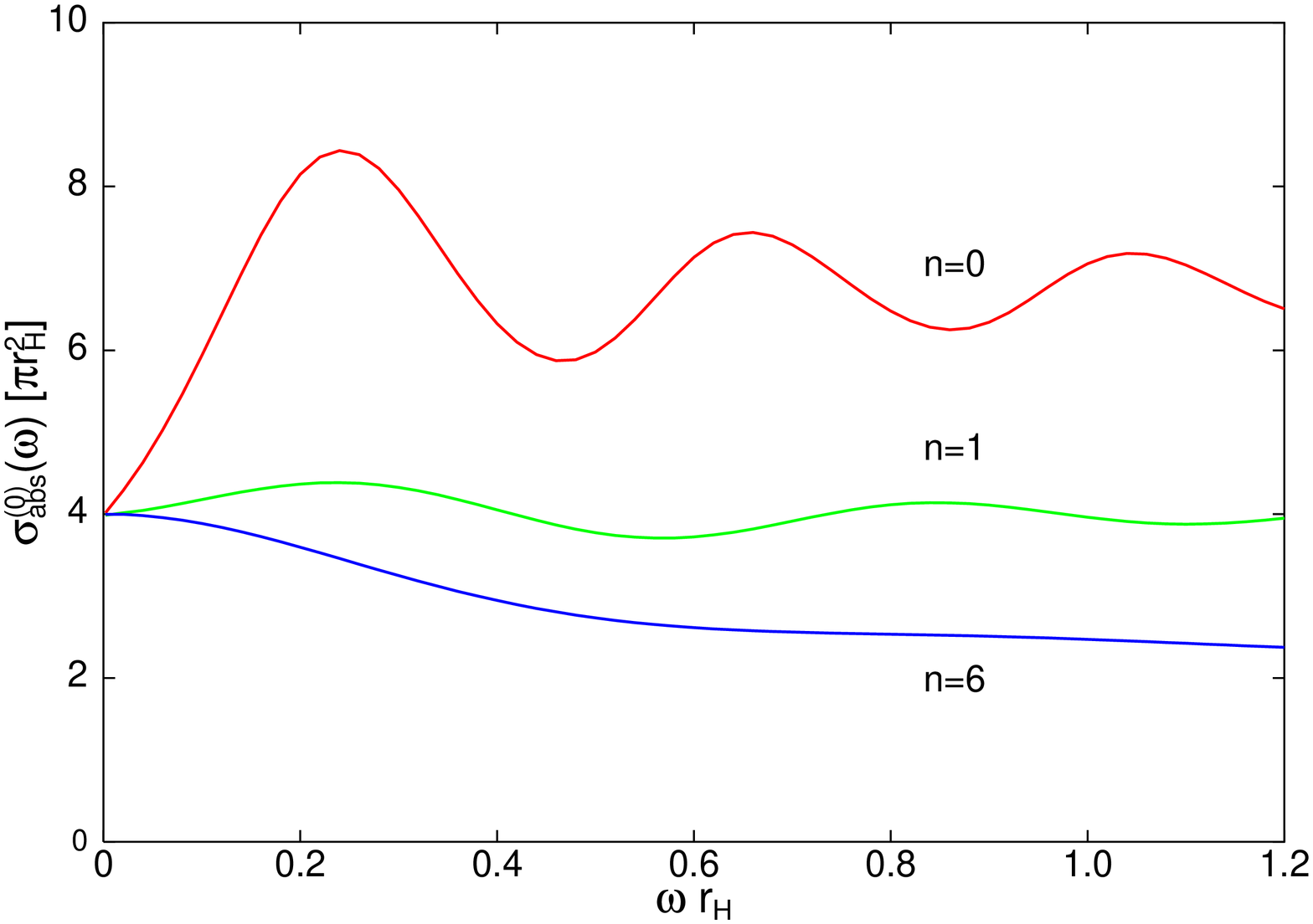}}}
\hspace*{4cm}}

\vspace*{-3.85cm}
\hspace*{5.5cm}\hbox{
\hspace*{0.4cm}\scalebox{0.55}
{\rotatebox{-90}{\includegraphics[width=7cm, height=10cm]{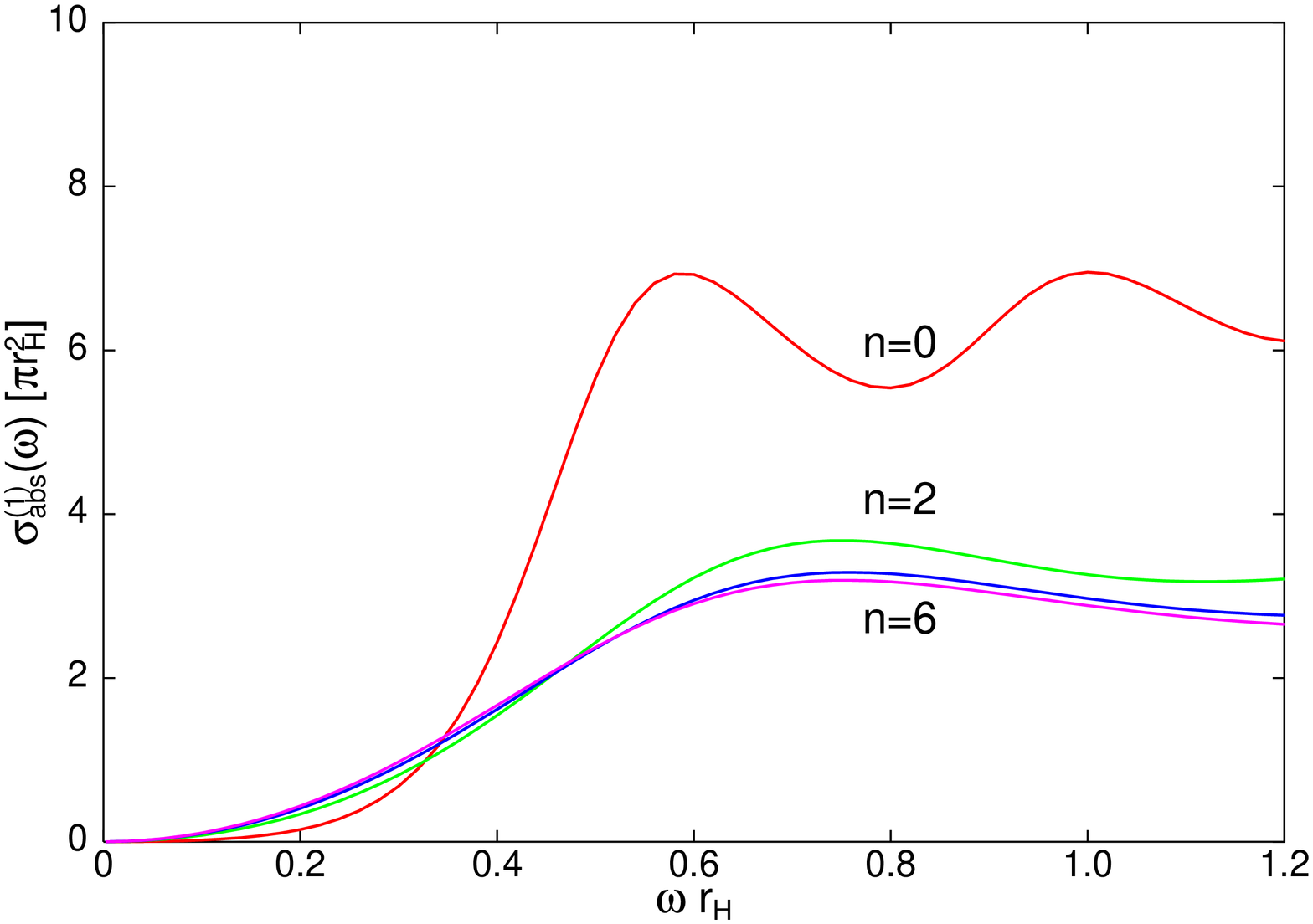}}}}
\caption{Absorption cross-section for brane-localised scalars (left plot) and
gauge bosons (right plot).}
\label{kansabs01}
\end{figure}

Whereas the Absorption Probability is a dimensionless quantity varying between
0 and 1 (in the non-rotating case), a dimensionful quantity may be constructed
out of it, namely the absorption cross-section, that is measured in units of
the horizon area ($\pi r_H^2$) and is defined as \cite{kanGKT}
\begin{equation}
\sigma_{\rm abs}(\omega)=\sum_{l}\frac{\pi r_H^2}{(\omega r_H)^2}\,(2l+1)
\,|{\cal A}(\omega)|^2\,.
\end{equation}
By following the approximate method, described above, to solve the radial
equation, one may compute the absorption probability and from that the
absorption cross-section. As an indicative case, in Fig. \ref{kansabsfer}(a)
we present the result for $\sigma_{\rm abs}(\omega)$ for the case of fermions
propagating in the projected-on-the-brane black-hole background. As we
observe, the horizontal axis does not extend to large values of the energy 
parameter $\omega r_H$; the reason for this is that, during the matching of
the two asymptotic solutions, the assumption was made that $\omega r_H \ll 1$,
that inevitably restricts the validity of the analytic result to small values
of the energy. Therefore, the behaviour of $\sigma_{\rm abs}(\omega)$, or
${\cal A}(\omega)$, for arbitrary values of the energy can only be derived
if numerical techniques are employed \cite{kanHK} for the solution of the
radial equation.
Then, the plot appearing in Fig. \ref{kansabsfer}(b) can be constructed. 
The qualitative agreement between the two plots is obvious and one can
see that the low-energy behaviour of $\sigma_{\rm abs}(\omega)$ is accurately
reproduced by the analytic result. However, as $\omega r_H$ increases, 
deviations start appearing. To complete the picture, in Figs. \ref{kansabs01}(a,b)
we present the behaviour of the absorption cross-section for scalars and
gauge bosons \cite{kanHK}, respectively. What is important in the behaviour of
$\sigma_{\rm abs}(\omega)$ is that {\bf (a)} it behaves differently for each
species of fields, and {\bf (b)} has a rather strong dependence on the
number of spacelike dimensions that exist transversely to the brane.

\begin{figure}[b]
\sidecaption
\includegraphics[scale=0.5]{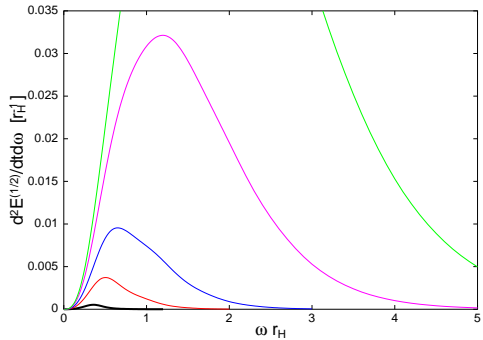}
\caption{Hawking radiation energy emission rates for brane-localised fermions,
for $n=0,1,2,4$ and $6$ (from bottom to top).}
\label{kanspectrum}
\end{figure}

When the (numerically) computed absorption probability and the temperature of
the black hole are substituted in the formula for the energy emission rate,
we obtain the radiation spectrum \cite{kanHK} that, for the indicative case
of fermions, is depicted in Fig. \ref{kanspectrum}. The different curves on
the plot stand for the differential energy emission rates per unit time and
unit frequency for the cases with
$n=0,1,2,4$ and 6 (from bottom to top). We may easily observe that the energy
emission rate is greatly enhanced by the number of extra spacelike dimensions,
a result that holds also for scalars and gauge bosons. In order to derive the
total emissivity, i.e. the energy emitted over the whole frequency regime per
unit time, we integrate over $\omega r_H$. The results for all species of 
brane-localised fields are presented in the first three rows of Table \ref{kanemissiv}.
From there, we may see that the total emissivity for the SM fields is enhanced
up to three or four orders of magnitude with the number of extra dimensions.

\begin{figure}[t]
\hspace*{0.0cm}\hbox{
\scalebox{0.55}{\rotatebox{0} 
{\includegraphics[width=10cm, height=7cm]{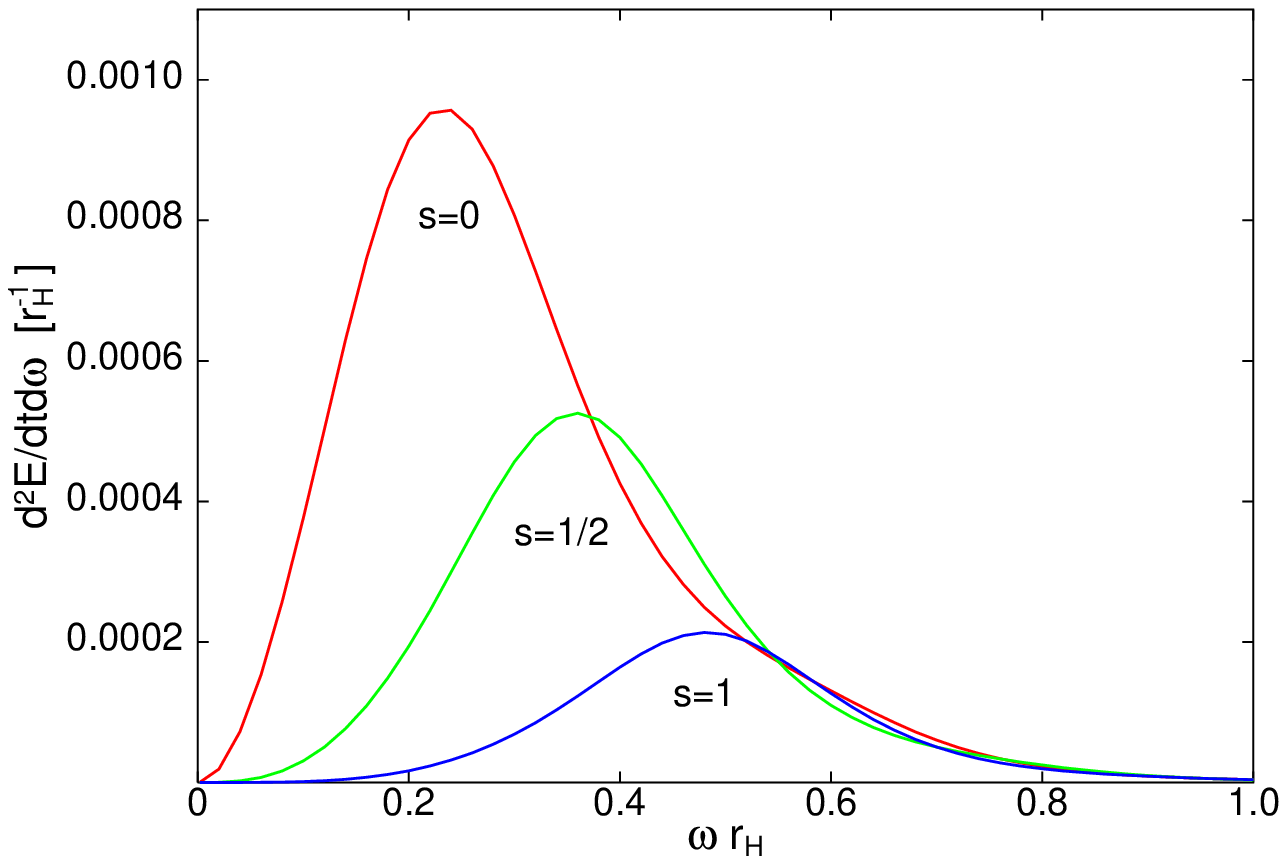}}}
\hspace*{0.4cm}\scalebox{0.55}
{\rotatebox{0}{\includegraphics[width=10cm, height=7cm]{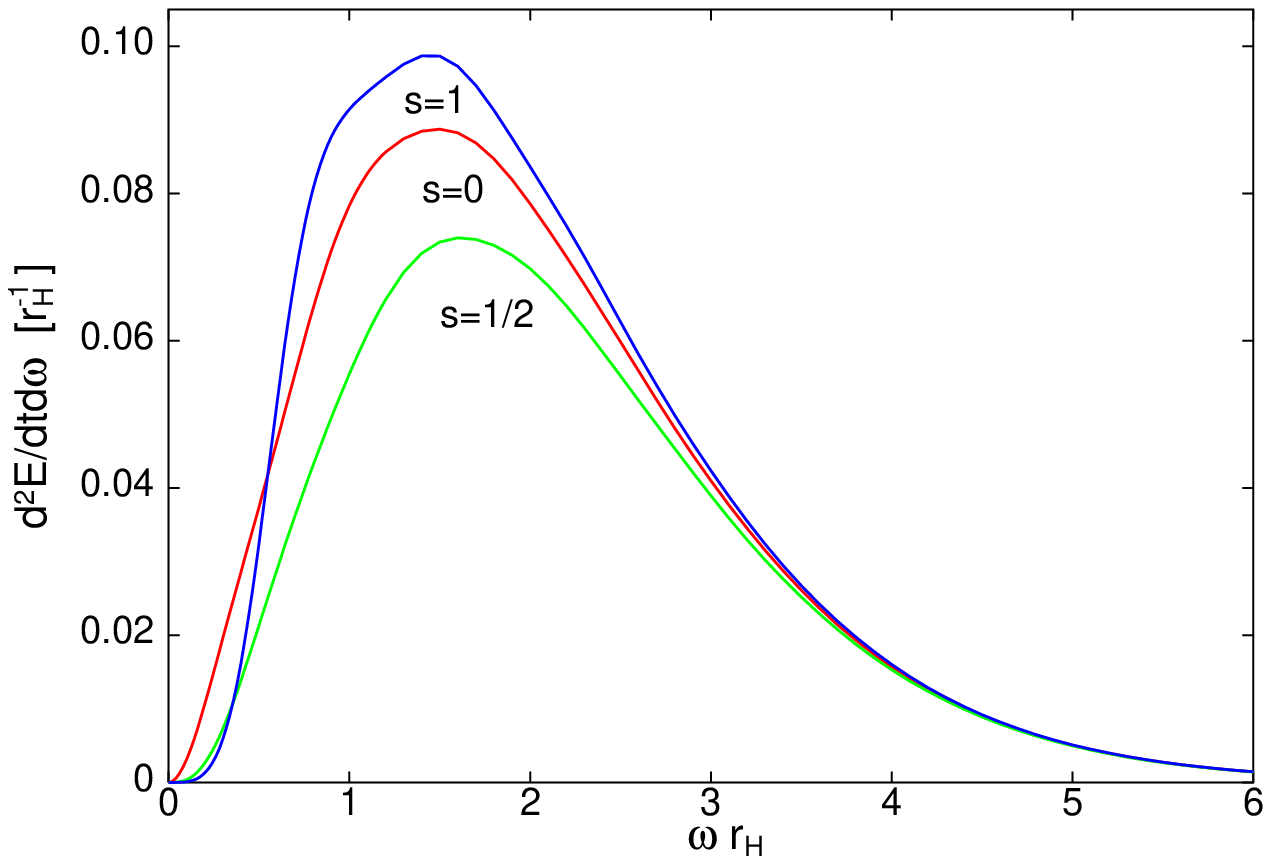}}}}
\caption{Relative emissivities for brane-localised fields for $n=0$ (left
plot) and $n=6$ (right plot).}
\label{kanrelative}
\end{figure}

We finish this subsection with an interesting observation that applies for
the relative emissivities of brane-localised fields. We have already seen
that the absorption cross-section, and consequently the absorption probability,
has a strong dependence on the spin of the propagating field. One thus expects
that different species of particles will have different emission rates. Indeed,
this may be seen by putting the emission curves of scalars, fermions and gauge
bosons on the same graph. For a purely 4-dimensional black hole \cite{kanPage},
this is shown in Fig. \ref{kanrelative}(a). According to this, the dominant
type of particles emitted by a black hole in 4 dimensions is scalars; then
come the fermions and finally the gauge bosons. Figure \ref{kanrelative}(b)
shows the same emission curves but in the case of a 10-dimensional black
hole. Here, the gauge bosons are the particles preferably emitted by the
black hole, then come the scalars and lastly the fermions. Therefore,
the number of extra dimensions determines not only the amount of energy
emitted per unit time by the black hole but also the type of the emitted
particles. 

\begin{table}[t]
\caption{Total emissivities for brane-localised scalars, fermions and
gauge bosons \cite{kanHK} and bulk gravitons \cite{kanCCG}}
\label{kanemissiv}
\begin{tabular}{ccccccccc} \hline\noalign{\smallskip}
\hspace*{0.5cm} $n$ \hspace*{0.5cm} & \hspace*{0.5cm} 0 \hspace*{0.4cm}  &
\hspace*{0.4cm} 1 \hspace*{0.4cm} & \hspace*{0.4cm} 2 \hspace*{0.4cm} & 
\hspace*{0.4cm} 3 \hspace*{0.4cm} & \hspace*{0.5cm} 4 \hspace*{0.5cm} &
\hspace*{0.5cm} 5 \hspace*{0.5cm} & \hspace*{0.5cm} 6 \hspace*{0.5cm} & 
\hspace*{0.5cm} 7  \hspace*{0.3cm}\\
\noalign{\smallskip}\svhline\noalign{\smallskip}  
Scalars  & 1.0 & 8.94 & 36.0 & 99.8 & 222 &
429 & 749 & 1220\\
 Fermions  & 1.0 & 14.2 & 59.5 & 162 & 352 &
664 & 1140 & 1830\\
G. Bosons  & 1.0 & 27.1 & 144 & 441 & 1020 &
2000 & 3530 & 5740 \\ 
Gravitons  & 1.0 & 103 & 1036 & 5121 & 2$\times$ 10$^4$ &
7$\times$ $10^4$ & 2.5$\times$ $10^5$ & 8$\times$ $10^5$\\
\noalign{\smallskip}\hline\noalign{\smallskip}
\end{tabular}
\end{table}
 
\subsection{The Spin-down Phase on the Brane}
\label{kanSpinphase}

We now turn to the phase in the life of the black hole that precedes the
Schwarzschild one. This is the spin-down phase during which the black hole
has a non-vanishing angular momentum -- this is the most generic situation
for a black hole created by a non-head-on particle collision. Assuming that
the produced black hole has an angular-momentum component only along an
axis in our 3-dimensional space, the line-element that describes the gravitational
background around such a higher-dimensional black hole is given by the
Myers-Perry solution \cite{kanMP}
\begin{eqnarray}
\hspace*{-1cm}
ds^2 &=& \left(1-\frac{\mu}{\Sigma\,r^{n-1}}\right)dt^2+\frac{2 a\mu\sin^2\theta}
{\Sigma\,r^{n-1}}\,dt\,d\varphi-\frac{\Sigma}{\Delta}dr^2 \nonumber \\[1mm] 
&-& \hspace*{0.3cm}
\Sigma\,d\theta^2-\left(r^2+a^2+\frac{a^2\mu\sin^2\theta}{\Sigma\,r^{n-1}}\right)
\sin^2\theta\,d\varphi^2 - r^2 \cos^2\theta\,d\Omega^2_n\,,
\label{kanMPmetric}
\end{eqnarray}
where
\begin{equation}
\Delta=r^2+a^2-\frac{\mu}{r^{n-1}}\,, \quad \quad\Sigma=r^2+a^2\cos^2\theta\,.
\end{equation}
The parameters $\mu$ and $a$ that appear in the metric tensor are associated to
the black hole mass and angular momentum, respectively, through the relations
\begin{equation}
M_{BH}=\frac{(n+2) A_{2+n}}{16 \pi G}\,\mu \qquad {\rm and} \qquad 
J=\frac{2}{n+2}\,a\,M_{BH}\,, \label{kan-mu}
\end{equation}
where $A_{2+n}$ is the area of a $(2+n)$-dimensional unit sphere. The horizon
radius is found by setting $\Delta(r_H)=0$ and is found to be:
$r_H^{n+1}=\mu/(1+a_*^2)$, where we have defined the quantity $a_* \equiv a/r_H$.
Finally, the temperature and rotation velocity of this black hole are given by
\begin{equation}
T_{H}=\frac{(n+1)+(n-1)\,a_*^2}{4\pi(1+a_*^2)\,r_{H}}\,,
\qquad \Omega_H=\frac{a}{(r_H^2+a^2)}\,. \label{kan-Temprot}
\end{equation}

Since we are still interested in the emission of brane-localised modes by the
black hole, we should first determine the line-element on the brane. As in the
case of the Schwarzschild phase, this will follow by fixing the values of the
``extra'' angular coordinates. This results in the disappearance of the
$d\Omega_{n}^{2}$ part of the metric leaving the remaining unaltered. Then,
by employing again the Newman-Penrose method, we compute the two -- decoupled
again -- master equations, one for the radial part of the field and one for the
angular part, namely \cite{kanreview, kanCKW}
\begin{equation}
\Delta ^{-s}\,\frac {d}{dr} \left(\Delta^{s+1}\,
\frac {d R_s}{dr}\right) + \left[\frac{K^2-iKs\Delta '}{\Delta}+4i s\omega r +
s\left( \Delta '' -2 \right)\delta_{s,|s|}-\Lambda^m_{sj}\right] R_s=0 
\label{kan-radialrot}
\end{equation}
and
\begin{eqnarray} && \hspace*{-1.0cm}\frac{1}{\sin\theta}\,\frac{\textstyle d}
{\textstyle d\theta}\left(\sin\theta\frac{\textstyle d S^m_{sj}}
{\textstyle d\theta}\right)+ \left[-\frac{\textstyle 2ms\cot\theta}{\textstyle \sin\theta}
-\frac{m^2}{\sin^2\theta}+a^2\omega^2\cos^2\theta\right. \nonumber \\[4mm]
&&\hspace*{3cm} \left.- 2 a \omega s \cos\theta +
s-s^2\cot^2\theta+\lambda_{sj}\right]S^m_{sj}(\theta)=0\,.
\label{kan-angulareq}
\end{eqnarray}
In the above, $S^m_{sj}(\theta)$ are the spin-weighted spheroidal harmonics
\cite{kanFlammer}, and we have used the definitions
\begin{equation}
K=(r^2+a^2)\,\omega-am\,, \qquad \Lambda^m_{sj}=\lambda_{sj} + a^2 \omega^2
-2 am \omega\,.
\end{equation}
The angular eigenvalue $\lambda_{sj}$ does not exist in closed form but it
may be computed either analytically, through a power series expansion in terms
of the parameter $a \omega$ of the form \cite{kanStaro,
kanFackerell, kanSeidel}
\begin{equation}
\lambda_{sj}=-s(s+1)+ \sum_{k}\,f^{j m s}_{k}\,(a \omega)^{k}\,,
\end{equation}
 or numerically \cite{kanHK2, kanDHKW, kanCKW, kanCDKW}.

\begin{figure}[t]
\mbox{\includegraphics[scale=0.18]{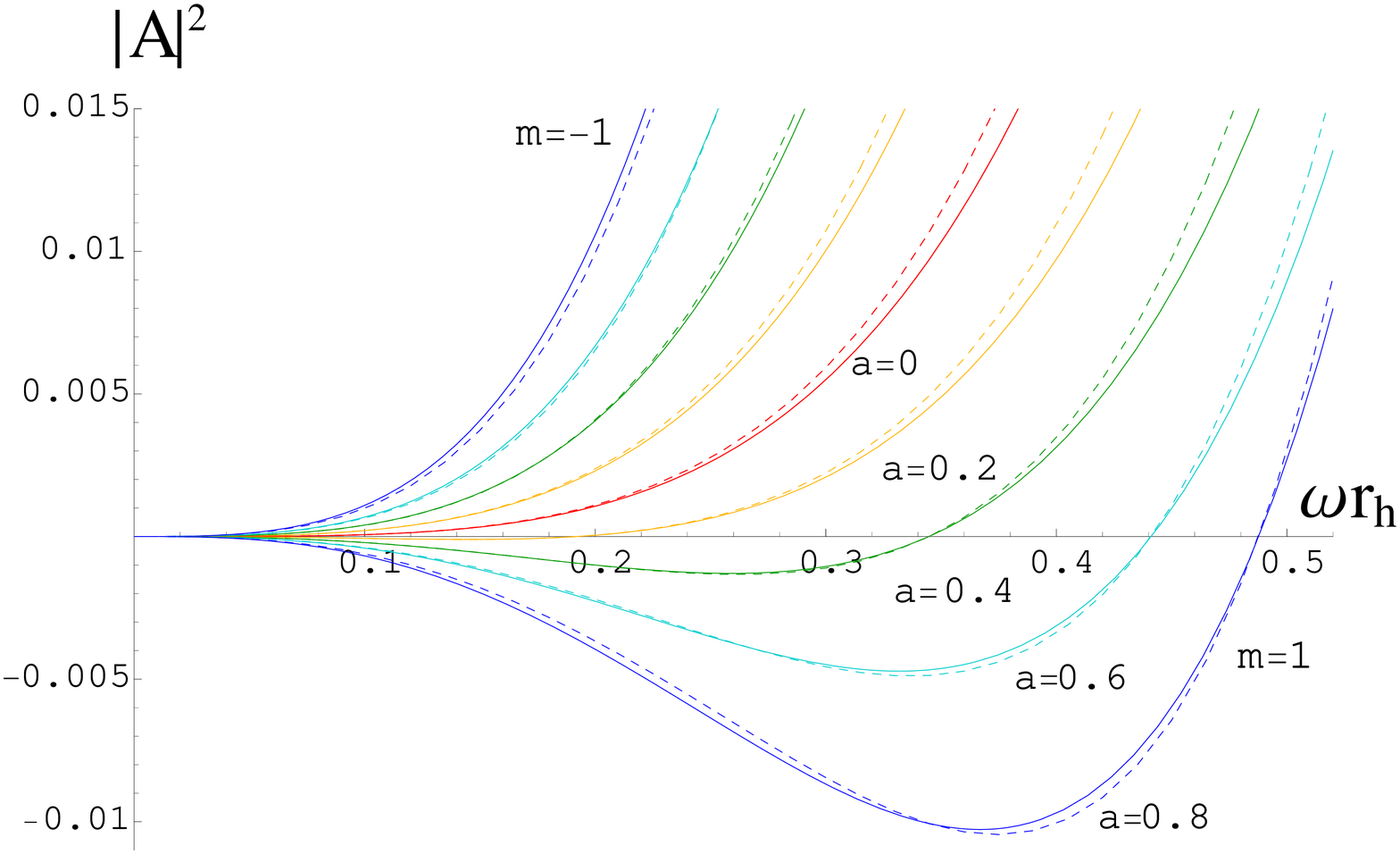}}
\includegraphics[scale=0.18]{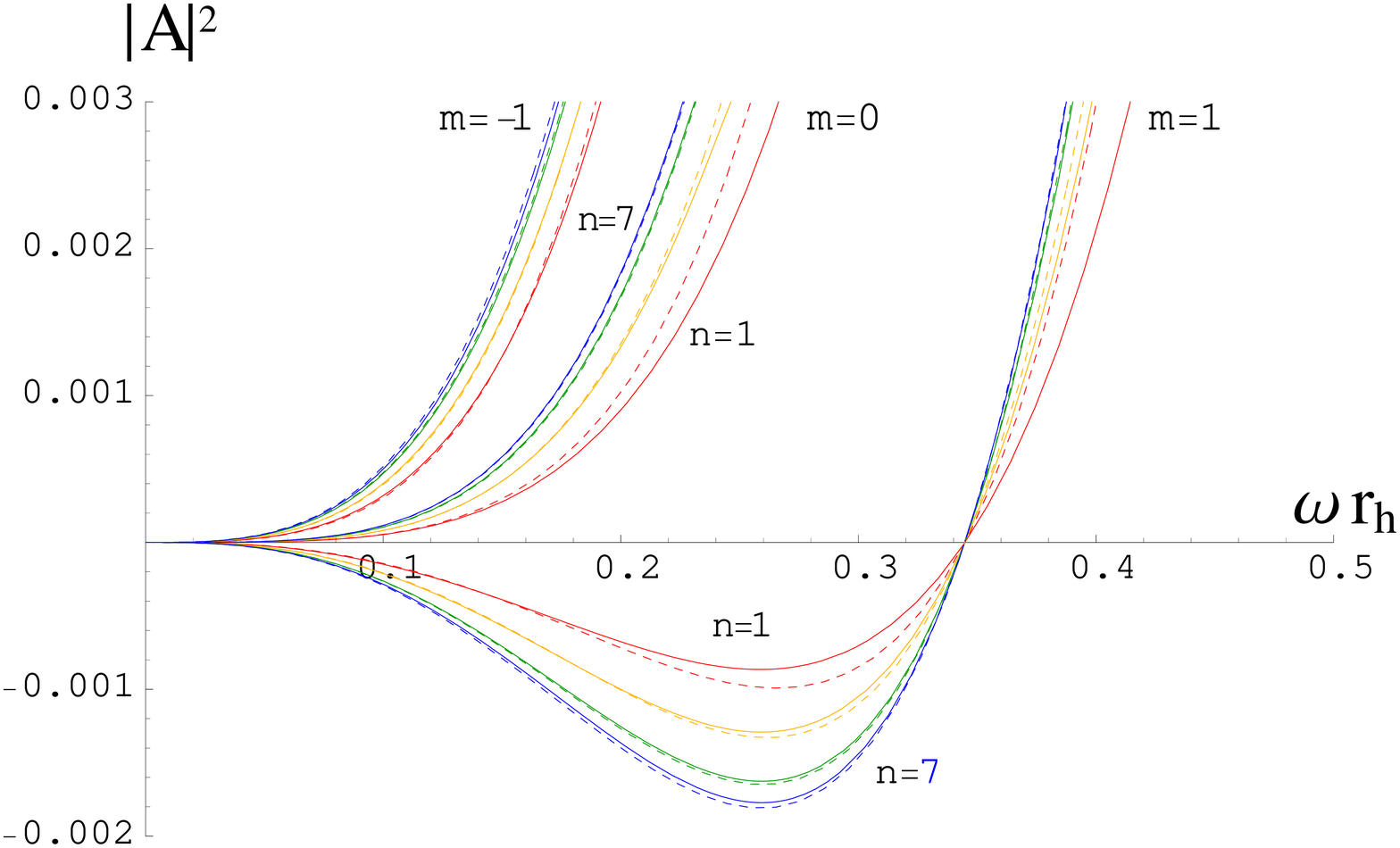}
\caption{Absorption probabilities for brane-localised scalar fields as a function
of the angular-momentum parameter $a$ (left plot) and number of extra dimensions
$n$ (right plot).}
\label{kan-Ascalar}
\end{figure}

The differential emission rates for the brane-localised modes during the spin-down
phase will be given by the 4-dimensional formula (\ref{kan-ratesrot}) but with the
greybody factor computed from the brane equation of motion (\ref{kan-radialrot})
and the temperature given by Eq. (\ref{kan-Temprot}). Despite the complexity of the
gravitational background, the absorption probability $|{\cal A}(\omega)|^2$
can be again found analytically in the low-energy and low-angular-momentum
regime. For example, in the case of scalar fields, the dependence of $|{\cal A}(\omega)|^2$
on the angular momentum parameter $a$ and number of extra dimensions $n$ is
given in Fig. \ref{kan-Ascalar} \cite{kanCEKT2}. Each curve in the two plots
is actually consisting of two lines: a solid one, representing our analytic
result, and a dotted one, representing the numerical result; it is clear that
in the low-$\omega$ regime, the agreement between the two sets of results is
indeed remarkable. A similar agreement is observed for the cases of fermions
and gauge bosons \cite{kanCEKT3}.

\begin{figure}[b]
\mbox{ 
\includegraphics[scale=0.45]{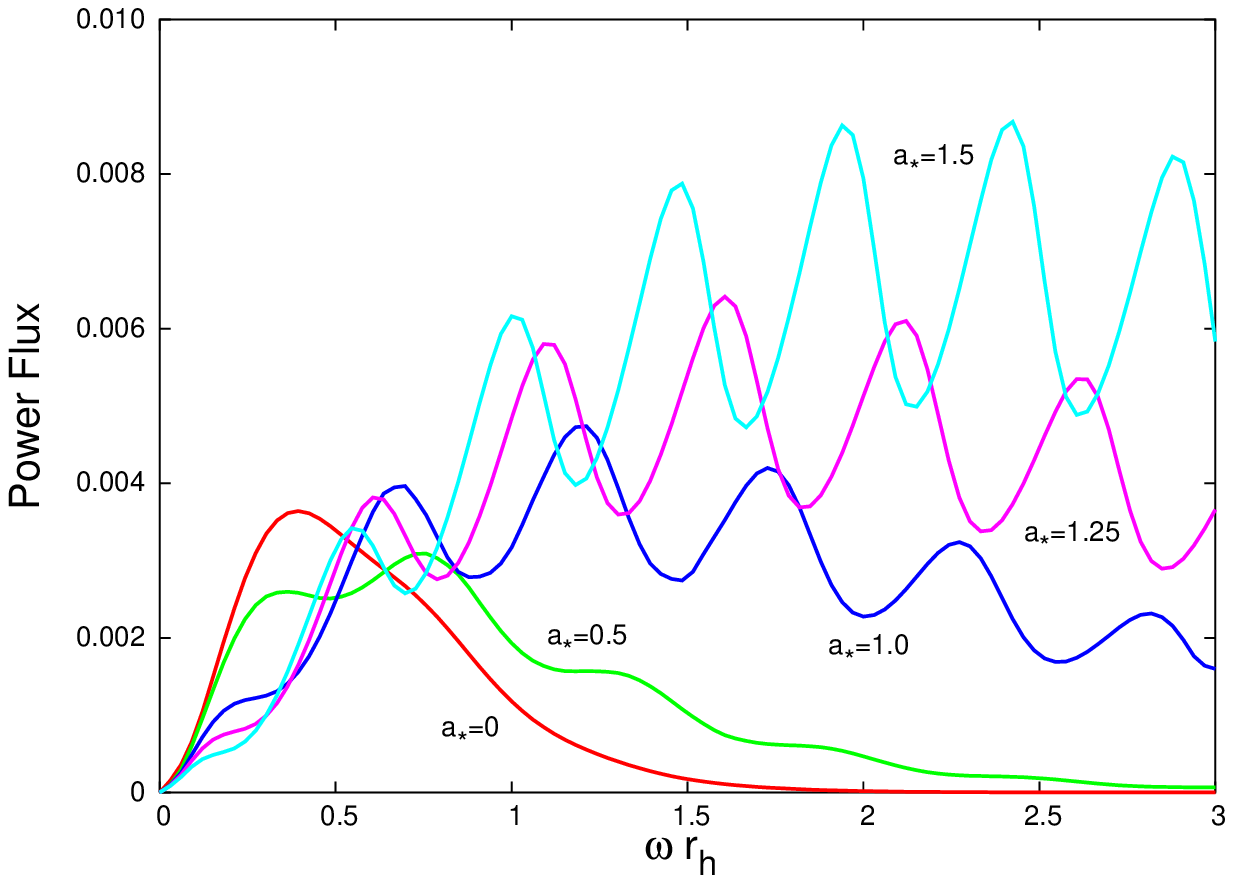}}
{\includegraphics[scale=0.45]{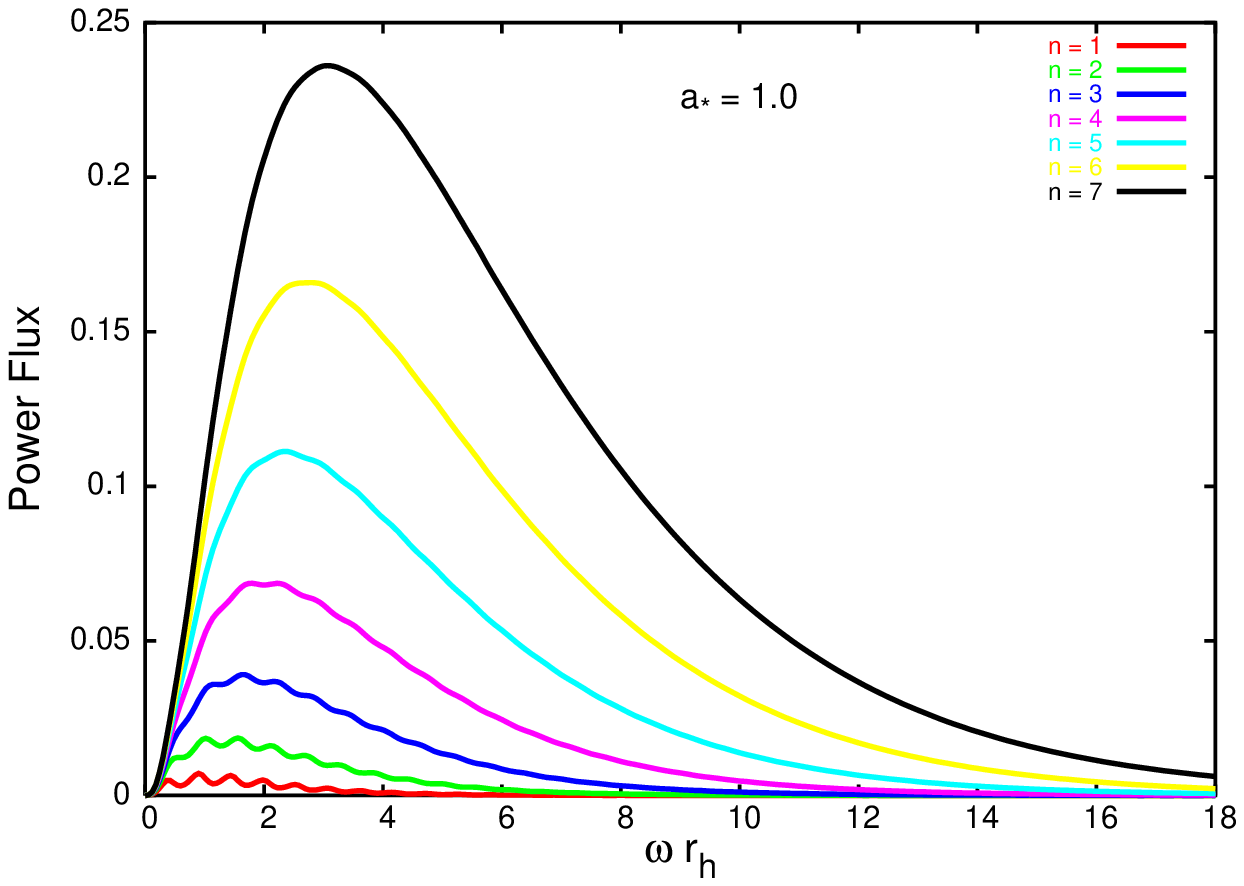}}
\caption{Energy emission rates for brane-localised scalar fields in terms of the
angular parameter (left plot) and gauge bosons in terms of the number of extra
dimensions (right plot).}
\label{kan-ratesrotplot}
\end{figure}

However, for the complete spectrum, we have to retort again to numerical 
analysis \cite{kanHK2, kanDHKW, kanIOP2, kanCKW, kanIOP3, kanCDKW}. In Fig.
\ref{kan-ratesrotplot}, we present the energy emission rates, for the
indicative cases of brane-localised scalars and gauge bosons, in terms
again of the angular-momentum parameter and number of extra dimensions.
It is clear that an increase in any of these two parameters results in the
significant enhancement of the energy emission rate. In Table \ref{kan-efrot},
we have put together the factors by which the energy emission rates are
enhanced, in terms of $a$ and $n$, for brane-localised scalars
\cite{kanHK2, kanDHKW}, gauge bosons \cite{kanCKW} and fermions \cite{kanCDKW}.
When the angular momentum parameter increases from 0 to 1, the energy emission
rates, for an 8-dimensional black hole, increases by a factor from 3 to 6, whereas,
for a black hole with a fixed angular momentum parameter $a_*=1$, the enhancement
factor is of the order of 50-100 when $n$ increases from 1 to 7. If we finally
compare the relative emissivities of different species of fields, then, once
again, it is the gauge bosons that a higher-dimensional rotating black hole
prefers to emit on the brane.

\begin{table}[t]
\caption{Enhancement factors for the energy emission rates in terms of the
angular momentum parameter and number of extra dimensions}
\label{kan-efrot}
\begin{tabular}{c|ccc|ccc}  \hline\noalign{\smallskip}
\hspace*{0.6cm}  & \hspace*{0.3cm} ($n=4$) \hspace*{0.2cm} & \hspace*{0.2cm}
$a_*=0$ \hspace*{0.2cm} &
 \hspace*{0.2cm} $a_*=1.0$ \hspace*{0.3cm} & \hspace*{0.3cm} ($a_*=1)$ \hspace*{0.2cm}
 & \hspace*{0.2cm} $n=1$ \hspace*{0.2cm} & \hspace*{0.3cm} $n=7$ \hspace*{0.5cm}  \\
\noalign{\smallskip}\svhline\noalign{\smallskip}   
\hspace*{0.4cm} Scalars  \hspace*{0.4cm} & & 1 & $\geq$ 3  & &1 & $\geq$ 100\\
Fermions  & & 1 & 6 && 1 & 99 \\ 
G. Bosons  & & 1 & $\geq$ 5 & &1 &  $\geq$ 50 \\ 
\noalign{\smallskip}\hline\noalign{\smallskip}
\end{tabular}
\end{table}

\begin{figure}[b]
\begin{tabular}{lll}
\includegraphics[scale=0.35]{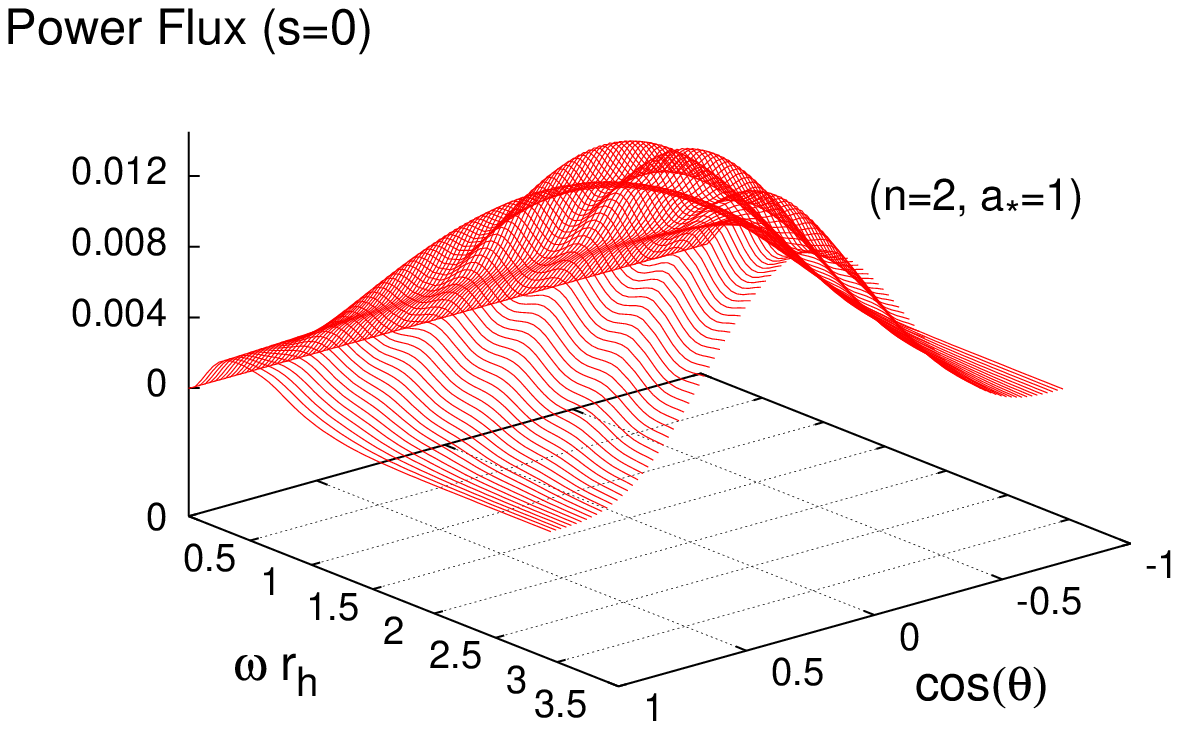} & \hspace*{-1cm}
\includegraphics[scale=0.35]{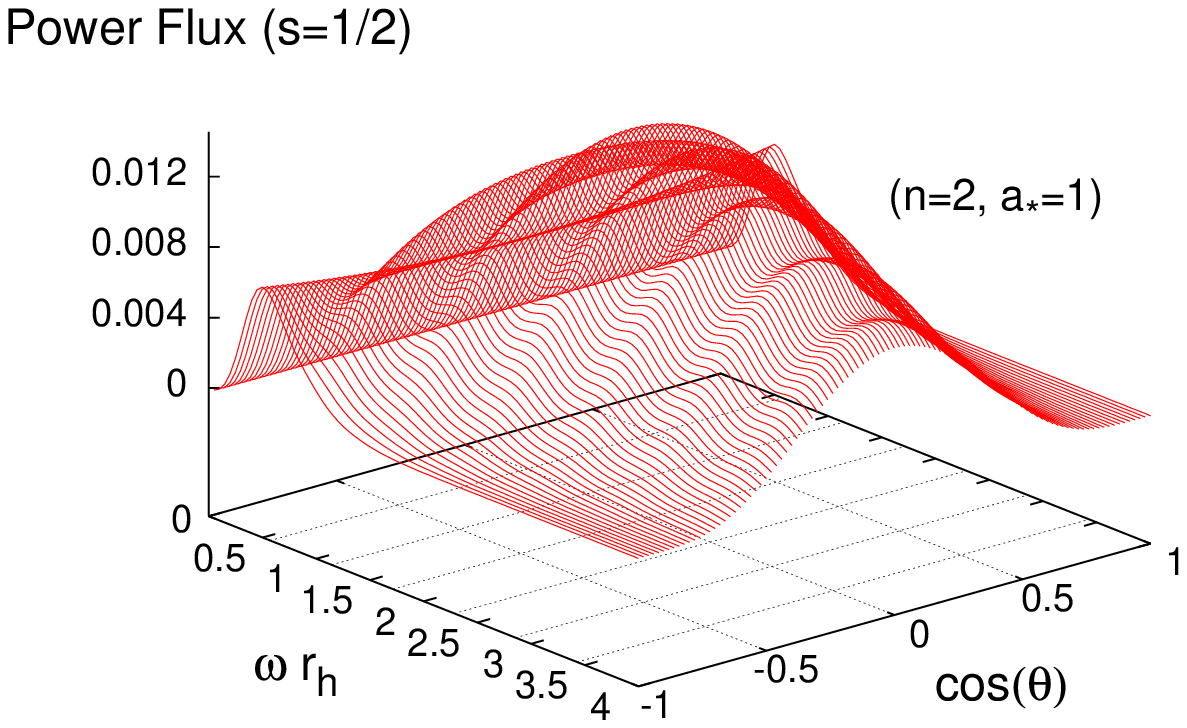} & \hspace*{-1cm}
\includegraphics[scale=0.35]{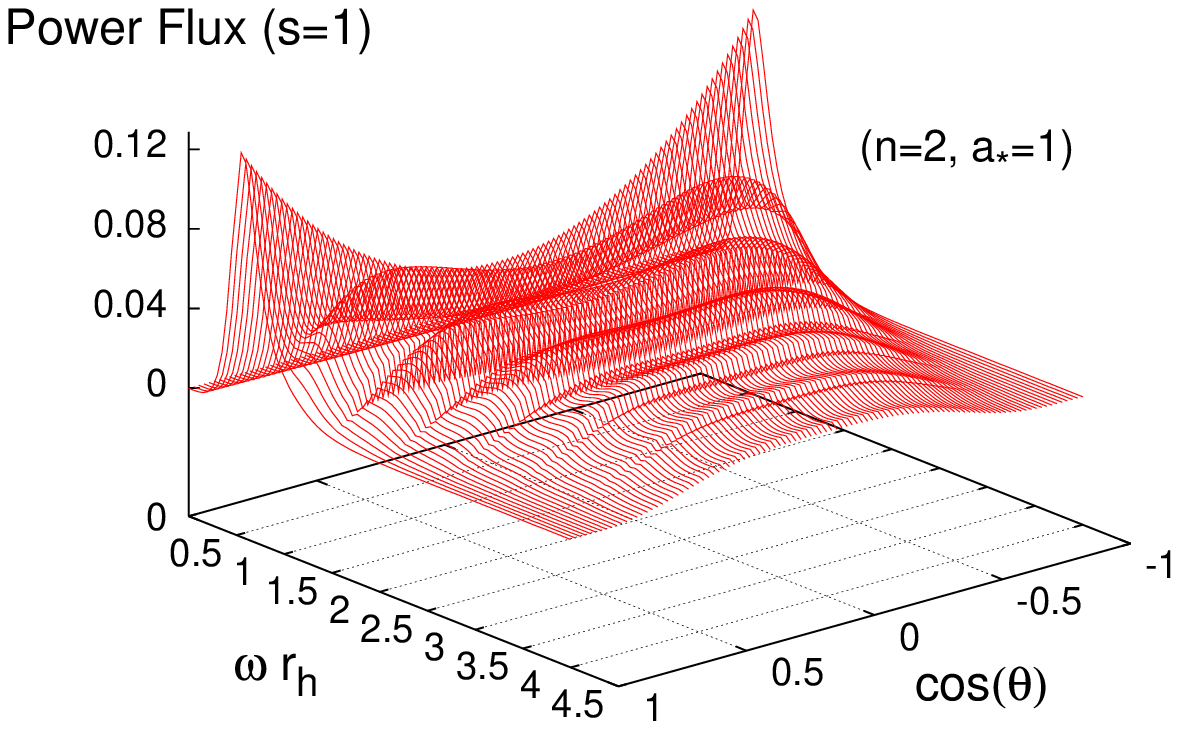} \end{tabular}
\caption{Angular distribution of the energy emission spectra for scalars
(left plot), fermions (central plot) and gauge bosons (right plot) for
a 6-dimensional black hole with $a_*=1$.}
\label{kan-angular}
\end{figure}

Let us finally comment on a particular feature that the radiation spectra
from the spin-down phase in the life of the black hole have. Unlike the
line-element that describes the background around the black hole during
its spherically-symmetric Schwarzschild phase, the one for the spin-down
phase possesses a preferred axis in space, that is the rotation axis of
the black hole. As a result, the radiation spectra of all emitted particles
have a non-trivial angular dependence. As an indicative case, in Fig.
\ref{kan-angular}, we present the energy emission rates for scalars,
fermions and gauge bosons, from a 6-dimensional, rotating black hole with
$a_*=1$, as a function of the energy parameter $\omega r_H$ and the $\cos(\theta)$
of the angle measured from the rotation axis of the black hole. In all
spectra, we observe that most of the energy is emitted along the equatorial
plane ($\theta=\pi/2$) as a result of the centrifugal force that is exerted
on all species of fields. In the special cases of fermions and gauge bosons,
i.e. of particles with non-vanishing spin, there is another effect, that
of the spin-rotation coupling, that causes an additional angular dependence
in their spectra, and aligns the emission along the rotation axis of the
black hole -- the effect is more dominant for gauge bosons than for
fermions, and it dies out as the energy of the emitted particles increases.
The angular spectra depicted in Fig. \ref{kan-angular} follow after solving
numerically the angular master equation (\ref{kan-angulareq}) for the value
of the spin-weighted spheroidal harmonics $S^m_{sj}(\theta)$ and calculating
the differential emission rate 
\begin{equation}
\frac {d^{3}E}{d(\cos\theta)dt  d\omega }  = 
\frac {1}{4\pi}\, \sum _{j =1}^{\infty }\,
\sum _{m=-j }^{j}
\frac {\omega \,\,|{\cal A}(\omega)|^2}{\exp\left(\tilde{\omega}/T_{H}\right) - 1}
\left[\left(S^m_{|s|j}\right)^2+\left(S^m_{-|s|j}\right)^2\right]\,,
\end{equation}
per unit time, frequency and solid angle for each species of particles
\cite{kanDHKW, kanCKW, kanCDKW}.

\subsection{Emission in the Bulk}

In the case that higher-dimensional mini black holes can indeed be created during
particle collisions, their detection becomes more likely if a significant part of
the black hole energy is channeled, through Hawking radiation emission, into brane
fields. Therefore, although the bulk emission will be interpreted as a missing energy
signal by a brane observer, we need to know the fraction of the total energy which
is lost along this channel. We thus need to study the emission by the black hole
of the species of particles that are allowed to propagate in the bulk, that is
gravitons and possibly scalar fields. The latter are the easier to study as their
equation of motion in the higher-dimensional spacetime can be easily found, by
generalizing its 4-dimensional expression, to be  
\begin{equation}
\frac{1}{\sqrt{-G}}\,\partial_M \left[\sqrt{-G}\,G^{MN}\,\partial_N \Phi\right]=0\,,
\end{equation}
where the capital indices take values in the range $(0,1,2,3,...,4+n)$ and $G_{MN}$
is the metric tensor of the higher-dimensional spacetime. 

\begin{figure}[t]
\sidecaption
\includegraphics[scale=0.55]{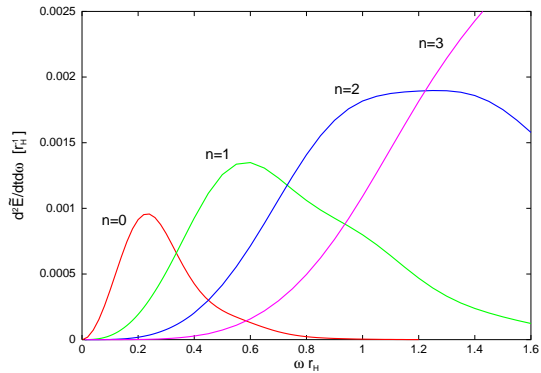}
\caption{Energy emission rates for bulk scalar fields, as a function of the number
of additional spacelike dimensions, for the Schwarzschild phase.}
\label{kanbulkS}
\end{figure}

We will study first the Schwarzschild phase, for which more results are available
in the literature. In that case, the gravitational background that we need to
consider is the higher-dimensional Schwarzschild-Tangherlini one (\ref{kanST}).
By assuming again a factorised ansatz for its wavefunction \cite{kanHK}
\begin{equation}
\Phi(t,r,\theta_i,\varphi)=e^{-i\omega t}\,R_{\omega l}(r)\,\tilde Y(\Omega)\,,
\end{equation}
where $\tilde Y(\Omega)$ is the higher-dimensional spherical harmonics \cite{kanMuller},
the equation of motion of the scalar field can reduce to a system of decoupled,
radial and angular, equations. From the radial one, we find the absorption probability
$|{\cal A}(\omega)|^2$ for a bulk scalar field, and finally the radiation spectrum
\cite{kanHK}. This is given in Fig. \ref{kanbulkS} in terms of the number of the
additional spacelike dimensions $n$. As in the case of brane emission, the energy
emission rate for bulk scalar fields is greatly enhanced as $n$ increases.

Therefore, the question ``which scalar channel, bulk or brane, is the most dominant
one?'' naturally arises. If the black hole has the choice to emit scalar fields both
on the brane and in the bulk, which channel is the most effective? In order to
answer this question, we need to compute the Bulk-to-Brane Relative Emissivity.
This follows by integrating the corresponding brane and bulk spectra for scalar
fields over the energy parameter $\omega r_H$, and computing their ratio. Then,
we obtain the values for the Bulk-to-Brane ratio displayed in Table \ref{kan-ratioS}
\cite{kanHK}. From these, we see that this ratio becomes smaller than unity as soon
as one extra dimension is introduced in the theory, decreases further as $n$ takes
intermediate values, and increases, while remaining smaller than unity, as $n$ 
reaches higher (supergravity-inspired) values. Thus, we deduce that, in general,
the brane scalar channel is the dominant one, however, for high values of $n$,
the bulk emission becomes indeed significant. 

\begin{table}[t]
\caption{Bulk-to-Brane Relative Emissivities Ratio for scalar fields in terms of $n$}
\label{kan-ratioS}
\begin{tabular}{c|cccccccc}  \hline\noalign{\smallskip}
 & \hspace*{0.4cm} $n=0$ \hspace*{0.2cm} & \hspace*{0.2cm} $n=1$ 
\hspace*{0.1cm} & \hspace*{0.1cm} $n=2$ \hspace*{0.1cm} & \hspace*{0.1cm} $n=3$
\hspace*{0.1cm} & \hspace*{0.1cm} $n=4$ \hspace*{0.1cm} & \hspace*{0.1cm} $n=5$ 
\hspace*{0.1cm} & \hspace*{0.1cm} $n=6$ \hspace*{0.1cm} & \hspace*{0.1cm} $n=7$
\hspace*{0.1cm} \\
\noalign{\smallskip}\svhline\noalign{\smallskip}  \hspace*{0.1cm} {\rm Bulk/Brane}
\hspace*{0.3cm}
& 1.0 & 0.40 & 0.24 & 0.22 & 0.24 & 0.33 & 0.52 & 0.93\\
\noalign{\smallskip}\hline\noalign{\smallskip}
\end{tabular}
\end{table}

The above result gives strong support to the argument presented in \cite{kanEHM}
where it was argued that most of the energy of a higher-dimensional black hole
will be emitted on the brane. The fact that the number of brane-localised
degrees of freedom is larger than the bulk ones, combined with the above
result that, when both channels are available, the black hole still prefers
the brane one, solidifies this argument. However, this matter is far from
settled since we have not looked yet at one of the most important species
of particles that may be emitted by the black hole into the higher-dimensional
spacetime, namely the gravitons. If the probability for graviton emission 
in the bulk comes out to be much higher than the one for lower-spin fields,
then, the bulk-to-brane balance may be overturned.

The graviton equation of motion in the bulk was derived \cite{kanKI} only a
few years ago, in the case of a spherically-symmetric, higher-dimensional
background. In there, a comprehensive analysis led to Schr\"odinger-like
equations for the three types of gravitational degrees of freedom that one
encounters in a higher-dimensional spacetime, namely tensor, vector and
scalar ones. In the years that followed, the equations of motion for all
three types were studied both analytically \cite{kanCNS, kanCEKT1} and
numerically \cite{kanCCG, kanJP}. The analytical approaches led to the
derivation of the gravitational radiation spectra either in the intermediate
\cite{kanCNS} or in the low-energy \cite{kanCEKT1} regime. 
In the latter case, it was shown \cite{kanCEKT1} that,
as long as the energy of the emitted particles remain in the lower part
of the spectrum, the total bulk graviton emission rate is sub-dominant to
the one for a bulk scalar field, which in turn is subdominant
to the one for a brane scalar field. However, a definite answer for the
graviton effect on the bulk-to-brane balance can be given only if the
complete spectrum for these degrees of freedom is known. This followed
from the numerical analysis performed in \cite{kanCCG}; according to their
results, the energy emission rates for gravitons for the Schwarzschild
phase of the black hole behave similarly to the ones for the other degrees
of freedom, i.e. they are significantly enhanced as the number of additional
spacelike dimensions increases. The exact enhancement factors in terms of
$n$ appear in the last row of Table \ref{kanemissiv}, and a direct comparison
is possible: clearly, the bulk graviton emission rate is the one that
exhibits the biggest in magnitude enhancement factor. 

\begin{table}[t]
\caption{Relative emissivities for brane-localised Standard Model fields
and bulk gravitons}
\label{kan-gravCCG}
\begin{tabular}{ccccccccc} \hline\noalign{\smallskip} 
$n$ & \hspace*{0.4cm} 0 \hspace*{0.4cm} & \hspace*{0.4cm} 1 \hspace*{0.4cm}
& \hspace*{0.4cm} 2 \hspace*{0.4cm} & \hspace*{0.4cm} 3\hspace*{0.4cm}  &
\hspace*{0.4cm}  4 \hspace*{0.4cm} & \hspace*{0.4cm} 5 \hspace*{0.4cm} &
\hspace*{0.4cm} 6 \hspace*{0.4cm} & \hspace*{0.4cm} 7\hspace*{0.4cm} \\
\noalign{\smallskip}\svhline\noalign{\smallskip} 
\hspace*{0.5cm} Scalars  \hspace*{0.5cm}& 1 & 1 & 1 & 1 & 1 &
1 & 1 & 1\\
 Fermions  & 0.55 & 0.87 & 0.91 & 0.89 & 0.87 &
0.85 & 0.84 & 0.82\\ 
G. Bosons  & 0.23 & 0.69 & 0.91 & 1.0 & 1.04 &
1.06 & 1.06 & 1.07 \\ 
Gravitons  & 0.053 & 0.61 & 1.5 & 2.7 & 4.8 &
8.8 & 17.7 & 34.7\\ 
\noalign{\smallskip}\hline\noalign{\smallskip}
\end{tabular}
\end{table}

Similar results follow when the relative emissivities are computed -- these
are displayed in Table \ref{kan-gravCCG}. Due to the aforementioned enhancement
factor, the gravitons, from an insignificant part of the total emission in 4
dimensions, become the dominant type of particles emitted by the black hole
as soon as $n \geq 2$. How does this affect the bulk-to-brane energy balance then?
Surprisingly, it is not in a position to overturn the dominance of the brane
channel. The reason for this is that in the relative emissivities for gravitons
displayed in Table \ref{kan-gravCCG} the total number of gravitational degrees
of freedom has already been taken into account. On the other hand, the relative
emissivities for the SM fields correspond to individual scalar, fermionic and
gauge bosonic degrees of freedoms. When the total number of SM degrees of freedom
(not to mention the beyond-the-SM ones) living on the brane is included in the
calculation of the total ``brane emissivity'', the brane channel turns out
to be the dominant one once again~\footnote{We note that in the presence of
higher-derivative curvature terms in the theory, such as the Gauss-Bonnet term,
it has been found \cite{kan-BGK} that the bulk emission might become the dominant
one for specific values of the black hole mass and Gauss-Bonnet coupling constant
even for the spherically-symmetric Schwarzschild phase. Also, in the case
that the model allows for fermions to propagate in the bulk, the bulk-to-brane
ratio in the fermionic channel exceeds unity even by an order of magnitude
\cite{kanCCDN}.}.

\begin{figure}[t]
\mbox{ 
\includegraphics[scale=0.75]{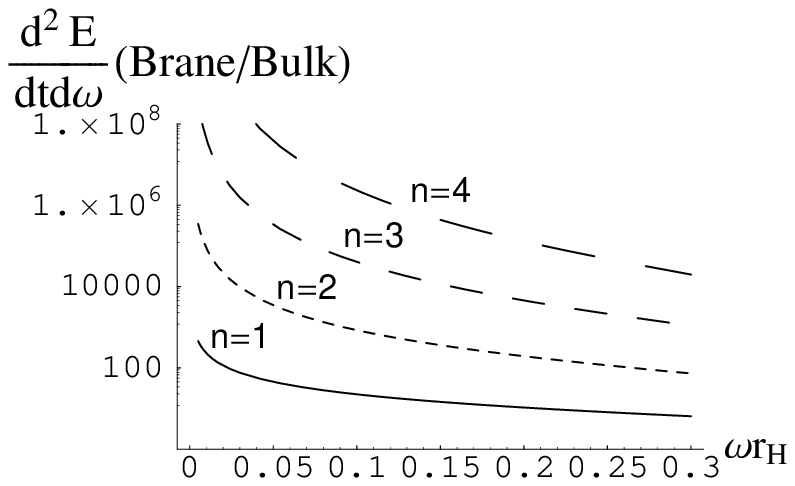}} \hspace*{-1.0cm}
{\includegraphics[scale=0.75]{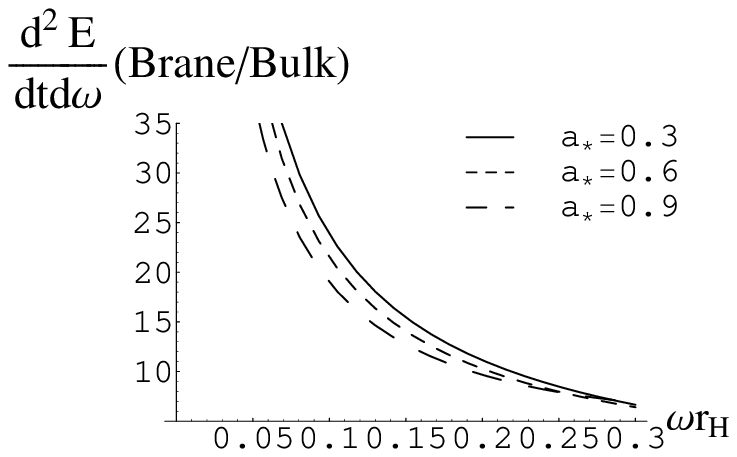}}
\caption{Brane-to-Bulk ratio of the differential energy emission rates for scalar
fields during the spin-down phase in terms of $n$ (left plot) and $a$ (right plot).}
\label{kan-bbrot}
\end{figure}

Have we therefore settled the question of the brane-to-bulk energy balance?
Perhaps, not. The discussion up to now referred to the Schwarzschild phase
in the life of the black hole, and another study needs to be performed for
the spin-down phase. The only results available in the literature for the
brane-to-bulk ratio in the case of a higher-dimensional, rotating black-hole
background, are the analytic ones for scalar fields presented in \cite{kanCEKT4}.
In Fig. \ref{kan-bbrot}, we display the ratio of the differential energy
emission rate for scalar fields living on the brane over the one for bulk
scalar fields from a higher-dimensional black hole with line-element given
by Eq. (\ref{kanMPmetric}).  From the left plot of Fig. \ref{kan-bbrot},
we see that for a black hole with fixed angular momentum ($a_*=0.5$) the
brane-to-bulk ratio remains above unity for all values of $n$. On the other
hand, from the right plot we observe that, for a 5-dimensional black hole,
the same ratio is again larger than unity but it decreases as either the
angular momentum of the black hole or the energy of the particle increases.
It would be indeed interesting to check whether the brane dominance in the
emission of scalar fields persists over the whole frequency regime, especially
for large values of $a$ \cite{kanCDKW2}.

\subsection{Deducing Basic Information }

Let us now discuss the methodology one should follow, in case we witness
the creation of miniature black holes in a collider experiment. In order to
deduce any useful information on the fundamental, higher-dimensional
theory, we need to compute with the greatest possible accuracy two quantities:
the mass of the black hole and its temperature. 

During a high-energy collision of composite particles, it is impossible
to know which pair of partons led to the creation of the black hole and what
was its total energy. Further losses of energy in the form of gravitational
or visible radiation during the balding phase complicates things even more.
The black-hole mass can therefore be reconstructed only through the 
measurement of the energy of the particles that appear in the final
state after the evaporation of the black hole \cite{kanDL}. Clearly, any
missing energy will greatly reduce the efficiency of the method, therefore
one needs to focus on events with little or no missing energy. To this
end, a cut is imposed on events with missing energy $E > 100$ GeV, so that
the black hole mass resolution is about 4\%, i.e. $\pm$200 GeV if
$M_{BH}=5$ TeV \cite{kanHPPRSW}.

The temperature of the black hole can be determined by performing a fit on
the detected Hawking radiation spectra \cite{kanDL}. Preferably, these spectra
should come from events involving only photons and electrons in the final state. 
The reason is that {\bf (a)} these events would have a very low background,
and {\bf (b)} the energy resolution of these particles is excellent even at
high energies. 

Once the temperature $T_H$ and mass $M_{BH}$ of the black hole are found with
the greatest possible accuracy, one could proceed to determine the dimensionality
of spacetime, in other words the value of $n$. From the temperature-horizon radius
relation (\ref{kantemp}), we may write \cite{kanDL}
\begin{equation}
\log (T_H) = -\frac{1}{n+1}\,\log (M_{BH}) + const. \label{kanMT}
\end{equation}
Then, the value of $n$ can simply follow by determining the slope of the
straight-line fit of the data relating $M_{BH}$ and $T_H$. The above method
is naturally not free of problems -- indicatively, we may mention the following:

\begin{itemize}
\item[{$\bullet$}] The resolution in the measurement of the black-hole mass 
$M_{BH}$ may not be good

\item[{$\bullet$}] The black hole temperature $T_H$ changes (increases) as a function
of time as the evaporation progresses

\item[{$\bullet$}] The multiplicity of particles in the final state of the
evaporation decreases for high values of $n$

\item[{$\bullet$}] Secondary particles that do not come directly from the
evaporating black hole may obscure the spectrum
\end{itemize}

\begin{figure}[t]
\includegraphics[scale=0.52]{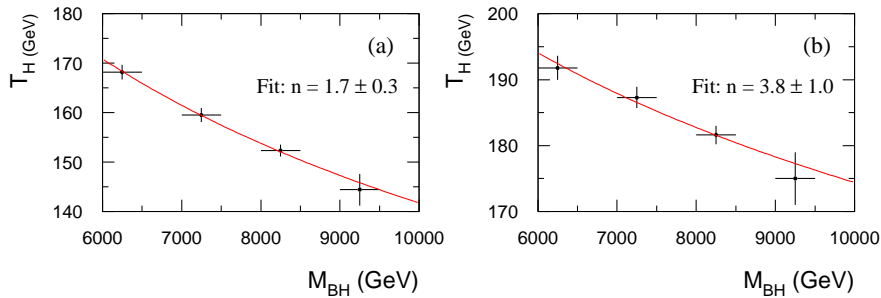}
\caption{Plots relating the black-hole mass and temperature measurements, and
the derived value of $n$, for constant (left plot) and variable (right plot)
temperature \cite{kanHPPRSW}.}
\label{kan-temp2cases}
\end{figure}

We have already discussed the first problem associated with the determination
of the black-hole mass. Let us briefly discuss the second one involving the
temperature. We consider the special case with fundamental Planck scale given
by $M_*=1$ TeV and number of additional dimensions $n=2$. We will pretend that
we do not know the value of $n$ but rather we are trying to find it through
Eq.  (\ref{kanMT}). We can assume that the temperature of the black hole
either remains constant or it increases as the time goes by. Then, the use
of Eq. ( \ref{kanMT}) leads to the two plots, respectively, appearing in
Fig. \ref{kan-temp2cases} \cite{kanHPPRSW}. As we see, by fitting the slope
of the straight line, we obtain $n=1.7 \pm 0.3$ in the first case, and
$n=3.8\pm1.0$ in the second.
A realistic model should be in a position to take into account that the
temperature of the black hole is indeed increasing as the evaporation progresses
but also that the lifetime of the black hole is extremely short. As a result,
the real situation should actually be somewhere in between the two cases
considered above, and an accurate fitting should be in a position to
produce the correct value of $n$ which lies indeed between the two derived values.

\begin{figure}[b]
\sidecaption
\includegraphics[scale=0.45]{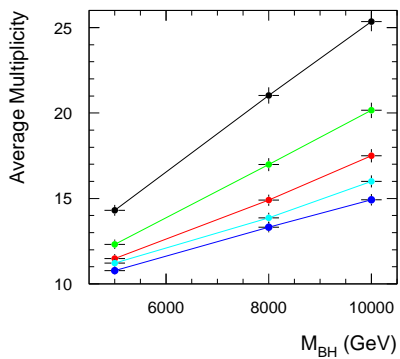}
\caption{Multiplicity of particles emitted by a black hole as the number of
the additional spacelike dimensions $n$ increases from 2 (top curve) to 6 
(bottom curve) \cite{kanHPPRSW}.}
\label{kanmultipli}
\end{figure}

The multiplicity of particles emitted by the evaporating black hole depends
strongly on the black hole mass and its temperature -- roughly, the first
quantity stands for the amount of energy available for emission and the 
second for the average energy that each emitted particle carries away. 
More accurately, the number of particles emitted by the black hole is given
by the relation \cite{kanDL}
\begin{equation}
\langle N \rangle = \langle \frac{M_{BH}}{E} \rangle \simeq 
\frac{M_{BH}}{2 T_H}\,.
\end{equation}
If, given the extremely short lifetime of the black hole, we assume that
its mass remains constant and that the black hole evaporates instantly into
a number of particles, the multiplicity then depends on the value of $T_H$.
From the entries of Table \ref{kanprop}, we see that the value
of the temperature increases as the number of additional dimensions $n$
increases, too. In Fig. \ref{kanmultipli}, we display a plot \cite{kanHPPRSW}
showing the multiplicity of particles emitted from a black hole as a function
of $M_{BH}$, and for various values of $n$ [increasing from 2 (top) to 6 (bottom)]
for $M_*=1$ TeV. From this, it is clear that, while for small values of $n$,
a black hole, that might be created at the LHC, can emit up to 25 particles,
for large values of $n$, this number drops at around 10. As a result, the
number of data points that we need to construct the $T_H-M_{BH}$ line
reduces significantly with $n$, and with it the accuracy in the determination of
its slope. While therefore, by using Eq. (\ref{kanMT}), we might be in a position
to obtain a rather accurate value of $n$ if that lies in the lower part of its
range, it might be very difficult to distinguish between the cases with $n=5$,
$n=6$ or $n=7$.

Many experiments, looking for beyond the SM physics, have included searches for
extra dimensions and miniature black holes in their research programs. At the
Large Hadron Collider alone, three collaborations (ALICE, ATLAS and CMS) are
planning to do so. But what type of particles and signatures should we expect
to see in the detectors? Will we be able to see the Hawking radiation emission
spectra that we presented in the previous sections for elementary SM degrees
of freedom (the so-called ``primary'' particles), or maybe ``secondary''
composite particles will be detected instead? In order to have a better
understanding of the type of particles expected to be seen in the final state,
we need a Black Hole Event Generator (BHEG) that simulates the black-hole
production and decay process given a number of initial conditions. The
method followed in a BHEG is roughly the following: 

\begin{itemize}
\item{} For a given center-of-mass energy $E$ of the colliding particles, the
black-hole mass $M_{BH}$ is estimated as a fraction of $E$
\item The theoretically predicted emission rates for the ``primary'' particles
are fed to the BHEG and the ``secondary'' particle spectra are produced
\end{itemize}

\begin{table}[t]
\caption{Predictions for the relative emissivities of SM fields \cite{kanHPPRSW}
derived by CHARYBDIS}
\label{kan-HERpred}
\begin{tabular}{ccccccccc} \hline\noalign{\smallskip} 
Type \hspace*{0.1cm} & \hspace*{0.1cm} Quarks  \hspace*{0.1cm} &
\hspace*{0.1cm}  Gluons \hspace*{0.1cm} & \hspace*{0.1cm} Charged leptons \hspace*{0.1cm}
& \hspace*{0.1cm}  Neutrinos \hspace*{0.1cm} & \hspace*{0.1cm} Photons \hspace*{0.1cm}
& \hspace*{0.1cm} $Z^0$ \hspace*{0.1cm} & \hspace*{0.1cm} $W^\pm$ \hspace*{0.1cm} &
\hspace*{0.1cm} Higgs\\
\noalign{\smallskip}\svhline\noalign{\smallskip}  
(\%) & 63.9 & 11.7 & 9.4 & 5.1 & 1.5 & 2.6 &
4.7 & 1.1 \\ 
\noalign{\smallskip}\hline\noalign{\smallskip}
\end{tabular}
\end{table}

At the moment, there are several Black Hole Event Generators that have been
constructed: CHARYBDIS \cite{kan-Char}, 
Catfish \cite{kan-Cat} and TRUENOIR \cite{kan-True}. For example, the
CHARYBDIS generator uses the HERWIG program \cite{kan-HER} to handle all the
QCD interactions, hadronization and secondary decays. It also makes specific
predictions for the relative emissivities of the different species of SM
particles expected to be detected. These are shown in Table \ref{kan-HERpred}
\cite{kanHPPRSW} from where we easily deduce that the dominant type of 
elementary particles emitted by the black hole should be the quarks.
The exact spectrum of emitted particles depends also on what happens during
the final phase in the life of the black hole, i.e. whether the black hole
evaporates completely by emitting a few energetic particles or a stable
remnant is formed \cite{kanACN, kanBDDO, kanBO, kanBOS, kanGidd, kanKBH}.
For this reason, BHEG's are equipped with an option regarding the nature
of the final state of the black hole that can be changed at will leading
each time to the corresponding radiation spectra. Finally, any observed
deviations from the anticipated behaviour stemming from standard QCD could
be considered as additional observable signatures of the black-hole formation.
For instance, QCD events with high transverse momentum are expected to become
gradually more rare as the energy of the collision increases \cite{kanBF};
on the contrary, black hole events with high transverse momentum dominate
over the QCD events with this happening at lower energies the smaller the
fundamental gravity scale $M_*$ is \cite{kanHKS}. In addition, in a standard
QCD process, one would expect to see the typical back-to-back di-jet production
seen in $p+p$ collisions with a particle distribution peaked at $\Delta\phi=0$
and $\pi$, with $\Delta\phi$ being the diference in the azimuthal coordinate between
the two emitted hadrons; as the black hole decays through the emission of individual,
sequential ``primary'' particles that lead to mono-jet events, we expect
the back-to-back di-jets to be strongly suppressed in the case of the black-hole
formation \cite{kanHKS, kanKBS}.

If we, therefore, wished to summarise some of the most interesting phenomena associated
with the existence of a low-scale gravity and the production of higher-dimensional
black holes, we should mention the following (see \cite{kanGidreview}, for a
complementary discussion on this):
\begin{itemize}
\item{} Large cross-sections, that increase with the center-of-mass energy of the
collision unlike every other SM process -- in such a case, the accurate measurement
of the cross-section could lead to the value of $M_*$

\item{} Primary particles emitted by the black hole with a thermal spectrum
and a much higher multiplicity than any other SM process

\item{} Energy emission rates and relative emissivities for different species of
fields determined by the number of additional spacelike dimensions

\item{} Non-trivial angular distribution in the radiation spectra coming from
the spin-down phase

\item{} Comparison of the observed with the predicted spectra could lead to the
detection of the final remnant -- its presence would increase the multiplicity of
particles in the final decay, lower the total transverse momentum by an amount
equal to its mass, and, if charged, it could even be directly detected via an
ionizing track in the detector

\item{} Events with high transverse momentum, above the expected QCD background

\item{} Strong suppression of back-to-back di-jet events contrary to the 
expected QCD behaviour

\item{} A significant amount of missing energy -- larger than the one for
SM or SUSY -- due to the emission of weakly interacting particles on the branes
and of gravitons or scalars in the bulk.
\end{itemize}


\subsection{Schwarzschild - de Sitter Black Holes}
\label{kan-SdS}

We would like to finish the discussion of the properties and fate of
higher-dimensional black holes with a brief reference to the class of
black holes that are formed in the presence of a positive cosmological
constant $\Lambda$ in the higher-dimensional spacetime. The geometrical
background around such a Schwarzschild - de Sitter black hole is given
by the line element
\cite{kanTangherlini}
\begin{equation}
ds^2 = - h(r)\,dt^2 + \frac{dr^2}{h(r)} + r^2 d\Omega_{2+n}^2\,,
\end{equation}
where
\begin{equation}
h(r) = 1-\frac{\mu}{r^{n+1}} - \frac{2 \kappa_D^2\,\Lambda\,r^2}{(n+3) (n+2)}\,,
\end{equation}
with $\kappa_D^2\equiv 8\pi G_D=8\pi/M_*^{2+n}$ and $\mu$ given again by
Eq. (\ref{kan-mu}). The equation $h(r)=0$ has two real, positive solutions, $r_H$
and $r_C$ standing for the black hole and the cosmological horizon, respectively.
The temperature of the black hole is given by the expression \cite{kan-KGB}
\begin{equation}
T_H = \frac{1}{\sqrt{h(r_0)}}\,\frac{1}{4\pi r_H}\,\Bigl[\,(n+1)- 
\frac{2\kappa^2_D \Lambda}{(n+2)}\,r_H^2\,\Bigl]\,,
\end{equation}
where $r_0$ is the value of the radial coordinate where the metric function
$h(r)$ reaches its maximum value - the presence of the factor $1/\sqrt{h(r_0)}$
in the expression for the temperature is necessary for its consistent definition
\cite{kan-HB}. A similar expression can be written for the temperature $T_C$
corresponding to the cosmological horizon -- the fact that $r_C>r_H$ guarantees
that $T_C<T_H$, therefore the flow of energy is from the black hole towards
the remaining spacetime.

\begin{figure}[t]
\hspace*{2cm}\includegraphics[scale=0.45]{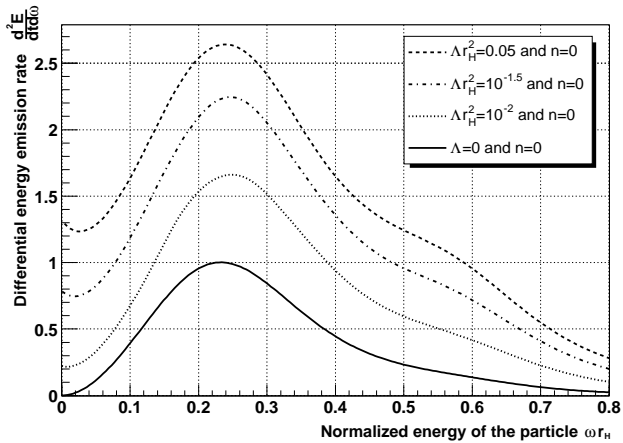}
\caption{Energy emission rate for a 4-dimensional black hole in terms of the
cosmological constant $\Lambda$.}
\label{kanLambda}
\end{figure}

The line-element of the gravitational background on the brane follows as
before by fixing the values of the additional angular coordinates to
$\theta_i=\pi/2$. It is then straightforward to write the equation of
motion of a scalar field propagating in the projected-to-the-brane background.
By solving the radial part of the equation of motion, we may again
determine the absorption probability $|{\cal A}(\omega)|^2$, and in
turn the energy emission rate. As in the case of a flat spacetime, the
energy emission rate is found to be greatly enhanced with the number of
extra dimensions $n$ but also with the value of the cosmological constant
$\Lambda$. This enhancement is clearly shown in Fig. \ref{kanLambda}
\cite{kan-KGB}. What is, however,
more important is the fact that, unlike in the case where $\Lambda=0$, 
for $\Lambda \neq 0$, the emission curve reaches an asymptotic non-zero value as
$\omega \rightarrow 0$. This asymptotic value increases with the value
of $\Lambda$ and it might, in principle, be used to ``read'' the value of
the cosmological constant from the observed radiation spectra. This non-zero
asymptotic value is due to the fact that, unlike in the case of a flat
spacetime, the absorption probability acquires a non-vanishing value
when $\omega \rightarrow 0$. This is given by the expression \cite{kan-KGB}
(see also \cite{kanHNS})
\begin{equation}
|{\cal A}(\omega=0)|^2=\frac{4r_C^2r_H^2}{(r_C^2+r_H^2)^2}\,,
\end{equation}
and is clearly caused by the presence of the cosmological horizon in the
theory -- in the limit $r_C \rightarrow \infty$, the asymptotic value of the
absorption probability, and in turn of the energy emission rate, reduces
to zero. This effect is independent of the existence of additional spacelike
dimensions and should be manifest also in the radiation spectra of 4-dimensional
primordial black holes. 


\section{Conclusions}
\label{kanconclusions}

In the context of the theories predicting the existence of either Large
or Warped Extra Dimensions, a low-scale gravitational theory, characterised
by a fundamental Planck scale $M_*$ much smaller than the 4-dimensional one
$M_P$, can be realised. This theory becomes accessible as soon as the
energy of a given experiment exceeds $M_*$, and manifests itself through
a number of strong gravity effects. These effects should be present even
at ordinary Standard Model particle collisions, taking place, for instance,
at ground-based colliders. As a matter of fact, it is expected that during 
collisions with $E>M_*$ we should witness the creation of not point-like
particles anymore but of extended heavy objects. One such type of objects
are black holes, one of the most fascinating classes of solutions in General
Relativity.

The Large Hadron Collider at CERN will have a center-of-mass energy of 14 TeV,
i.e. more than an order of magnitude larger than the value of the fundamental
Planck scale $M_*=1$ TeV, suggested by the most optimistic scenaria with
extra dimensions. It becomes then a natural place to look for strong gravity
effects, and possibly for the creation of black holes. Studies have shown
that their production can be realised as long as the energy of the collision
exceeds at least the value of 8 TeV. At the same time, the produced black
holes are expected to have a mass of at least a few times the value of the
fundamental Planck scale if we want the classical theory of General Relativity
and its predictions to be still applicable. According to the above restrictions,
the Large Hadron Collider is found to lie on the edge of both the classical
regime and of the black hole creation threshold.

The calculation of the value of the corresponding production cross-section
has attracted a great attention over the years. The current results seem to
support the claim that this value is significant and that it will lead to the
creation of, at least, a few black-hole events per day. In addition, the
study of the properties of these higher-dimensional black holes suggest that
the presence of extra dimensions greatly facilitates their creation: for
instance, the horizon radius of these black holes, although tiny, is orders
of magnitude larger compared to the one for a 4-dimensional black hole with
the same mass. 

When it comes to the detection of these events, the terms are also favourable.
The most important observable associated with the creation of the black hole
will be the emission of Hawking radiation, in the form of elementary particles,
as the black hole evaporates. The corresponding radiation spectrum will be
centered around the value of the temperature of the black hole, which, for 
the mass values that would allow their creation at LHC, comes out to be in
the range of ~100-600 GeV. A thorough theoretical study, employing either
analytic or numerical techniques, is necessary in order to determine, and
thus predict, the exact Hawking radiation spectrum from a decaying black hole.
The differential energy emission rates are found to depend on a number of
particle and spacetime properties, and thus to encode a valuable amount of
information for the gravitational background and for the species of particles
emitted. Some of the quantities on which we may deduce information are the
number of additional dimensions that exist transversely to the brane, the
black-hole angular momentum, the cosmological constant, the spin of the
emitted particles, and so on. 

In order to make a realistic prediction of the radiation spectra and also to
model in more detail the dynamical aspects of the production and evaporation
process, Black Hole Event Generators have been constructed. The exact form
of the radiation spectra, together with an additional number of distinct 
observable signatures should make the detection of Black Holes, and thus of
the existence itself of the extra dimensions, possible at the Large Hadron
Collider. Hopefully, during the coming years, our understanding of particle and
gravitational physics, and of the fundamental theory that describes them,
will be considerably extended beyond the current limits.

\begin{acknowledgement}
I am grateful to my collaborators (J. March-Russell, C. Harris, A. Barrau,
J. Grain, G. Duffy, E. Winstanley, M. Casals, S. Creek, O. Efthimiou,
K. Tamvakis and S. Dolan, in chronological order) for their valuable
help and inspiration while trying to uncover some of the secrets of the
higher-dimensional black holes. I would also like to thank the organisers
of the 4th Aegean Summer School on Black holes for their kind invitation
to present these lectures. I finally acknowledge financial support from
the UK PPARC Research Grant PPA/A/S/2002/00350 and participation in the
RTN Universenet (MRTN-CT-2006035863-1 and MRTN-CT-2004-503369).
\end{acknowledgement}


\begin{thebibliography}{99.}%

\bibitem{kanAntoniadis} I.~Antoniadis, 
{Phys. Lett.} B {\bf 246}, 377 (1990). 

\bibitem{kanHW} P.~Horava and E.~Witten,
Nucl.\ Phys.\  B {\bf 460}, 506 (1996); 
Nucl.\ Phys.\  B {\bf 475}, 94 (1996). 

\bibitem{kanLykken} J.~Lykken, 
{Phys. Rev.} D {\bf 54}, 3693 (1996). 

\bibitem{kanAkama}
K.~Akama, 
Lect.\ Notes Phys.  {\bf 176}, 267 (1982).

\bibitem{kanRuS} V.~A.~Rubakov and M.~E.~Shaposhnikov,
{Phys. Lett.} B {\bf 125}, 139 (1983); 
{Phys. Lett.} B {\bf 125}, 136 (1983). 

\bibitem{kanVisser} M.~Visser, 
{Phys. Lett.} B {\bf 159}, 22 (1985). 

\bibitem{kanWiltshire} G.~W.~Gibbons and D.~L.~Wiltshire,
{Nucl. Phys.} B {\bf 287}, 717 (1987). 

\bibitem{kanADD} N.~Arkani-Hamed, S.~Dimopoulos and G.~Dvali,
Phys. Lett. B {\bf 429}, 263 (1998); 
Phys. Rev. D {\textbf 59}, 086004 (1999). 

\bibitem{kanAADD} I.~Antoniadis, N.~Arkani-Hamed, S.~Dimopoulos and G.~R.~Dvali,
Phys. Lett. B {\textbf 436}, 257 (1998). 

\bibitem{kanRS} L.~Randall and R.~Sundrum,
Phys. Rev. Lett. {\textbf 83}, 3370 (1999); 
Phys. Rev. Lett. {\bf 83}, 4690 (1999). 

\bibitem{kanDelphi} P.~Abreu {\it et al.}  [DELPHI Collaboration],
{\it Eur. Phys. J.} {\bf C17}, 53 (2000). 

\bibitem{kanOpal} G.~Abbiendi {\it et al.}  [OPAL Collaboration],
{\it Eur. Phys. J.} {\bf C18}, 253 (2000). 

\bibitem{kanCDF} D.~Acosta {\it et al.}  [CDF Collaboration],
{\it Phys. Rev. Lett.} {\bf 89}, 281801 (2002). 

\bibitem{kanHoyle}
C.~D.~Hoyle, U.~Schmidt, B.~R.~Heckel, E.~G.~Adelberger, J.~H.~Gundlach,
D.~J.~Kapner and H.~E.~Swanson,
{\it Phys. Rev. Lett.} {\bf 86}, 1418 (2001). 

\bibitem{kanKapner}
D.~J.~Kapner, T.~S.~Cook, E.~G.~Adelberger, J.~H.~Gundlach, B.~R.~Heckel,
C.~D.~Hoyle and H.~E.~Swanson,
Phys.\ Rev.\ Lett.\  {\bf 98}, 021101 (2007). 

\bibitem{kanHall} L.~J.~Hall and D.~R.~Smith,
{\it Phys. Rev.} {\bf D60}, 085008 (1999). 

\bibitem{kanCullen} S.~Cullen and M.~Perelstein,
{\it Phys. Rev. Lett.} {\bf 83}, 268 (1999).  

\bibitem{kanBarger} V.~D.~Barger, T.~Han, C.~Kao and R.~J.~Zhang,
{\it Phys. Lett.} {\bf B461}, 34 (1999). 

\bibitem{kanReddy} C.~Hanhart, D.~Phillips, S.~Reddy and M.~Savage,
{\it Nucl. Phys.} {\bf B595}, 335 (2001). 

\bibitem{kanReddy2} C.~Hanhart, J.~A.~Pons, D.~R.~Phillips and S.~Reddy,
{\it Phys. Lett.} {\bf B509}, 1 (2001). 

\bibitem{kanPospelov} R.~Allahverdi, C.~Bird, S.~Groot Nibbelink and M.~Pospelov,
{\it Phys. Rev.} {\bf D69}, 045004 (2004). 

\bibitem{kanHR1} S.~Hannestad and G.~Raffelt,
{\it Phys. Rev. Lett.} {\bf 87}, 051301 (2001). 

\bibitem{kanHann} S.~Hannestad,
{\it Phys. Rev.} {\bf D64}, 023515 (2001). 

\bibitem{kanHR2} S.~Hannestad and G.~Raffelt,
{\it Phys. Rev. Lett.} {\bf 88}, 071301 (2002); 
{\it Phys. Rev.} {\bf D67}, 125008 (2003) [Erratum-ibid. {\bf D69}, 029901
(2004)]. 

\bibitem{kanGSS} M. Gogberashvili, A.S. Sakharov and E.K.G. Sarkisyan,
{\it Phys. Lett.} {\bf B644}, 179 (2007). 


\bibitem{kanFeng} L.~A.~Anchordoqui, J.~L.~Feng, H.~Goldberg and A.~D.~Shapere,
{\it Phys. Rev.} {\bf D68}, 104025 (2003).

\bibitem{kanBF}
T.~Banks and W.~Fischler,
hep-th/9906038. 

\bibitem{kanThorne}
K.~S.~Thorne, in {\it Magic without Magic}, ed. J.~R.~Klauder (San Fransisco,
1972).

\bibitem{kanNewHoop} D.~Ida and K.~i.~Nakao,
Phys.\ Rev.\  D {\bf 66}, 064026 (2002).

\bibitem{kanTangherlini} F.~R.~Tangherlini,
Nuovo Cim.\  {\bf 27}, 636 (1963). 

\bibitem{kanMP} R.~C.~Myers and M.~J.~Perry,
Annals Phys. {\bf 172}, 304 (1986). 

\bibitem{kanMR} P.~Meade and L.~Randall,
{\it JHEP} {\bf 0805}, 003 (2008).

\bibitem{kankhan}
K.A. Khan and R. Penrose, {\it Nature} {\bf 229}, 185 (1971).

\bibitem{kanszekeres}
P.~Szekeres, 
{\it J.\ Math.\ Phys.}  {\bf 13}, 286 (1972).

\bibitem{kandray}
T.~Dray and G.~'t Hooft,
{\it Nucl.\ Phys.} {\bf B253}, 173 (1985); 
{\it Class.\ Quant.\ Grav.}  {\bf 3}, 825 (1986). 

\bibitem{kanyurtsever}
U.~Yurtsever,
{\it Phys.\ Rev.} {\bf D38}, 1706 (1988); 
{\it Phys.\ Rev.} {\bf D38}, 1731 (1988). 

\bibitem{kanDP}
P.~D.~D'Eath and P.~N.~Payne,
{\it Phys.\ Rev.} {\bf D46}, 658 (1992); 
{\it Phys.\ Rev.} {\bf D46}, 675 (1992); 
{\it Phys.\ Rev.} {\bf D46}, 694 (1992). 

\bibitem{kanthooft}
G.~'t Hooft, 
{\it Phys.\ Lett.} {\bf B198}, 61 (1987); 
{\it Nucl.\ Phys.} {\bf B304}, 867 (1988); 
{\it Nucl.\ Phys.} {\bf B335}, 138 (1990). 

\bibitem{kanGS} S.~B.~Giddings and M.~Srednicki,
Phys.\ Rev.\  D {\bf 77}, 085025 (2008).

\bibitem{kangross}
D.~J.~Gross and P.~F.~Mende,
{\it Phys.\ Lett.} {\bf B197}, 129 (1987); 
{\it Nucl.\ Phys.} {\bf B303}, 407 (1988). 

\bibitem{kanamati}
D.~Amati, M.~Ciafaloni and G.~Veneziano,
{\it Phys.\ Lett.} {\bf B197}, 81 (1987); 
{\it Int.\ J.\ Mod.\ Phys.} {\bf A3}, 1615 (1988); 
{\it Phys.\ Lett.} {\bf B216}, 41 (1989). 

\bibitem{kanGGM} S.~B.~Giddings, D.~J.~Gross and A.~Maharana,
Phys.\ Rev.\  D {\bf 77}, 046001 (2008).

\bibitem{kanverlinde}
H.~Verlinde and E.~Verlinde, 
{\it Nucl.\ Phys.} {\bf B371}, 246 (1992). 

\bibitem{kanAS} P.C. Aichelburg and R.U. Sexl,
{\it Gen. Rel. Grav.} {\bf 2}, 303 (1971). 

\bibitem{kanGeneral}   V.~P.~Frolov and I.~D.~Novikov,
{\it ``Black hole physics: Basic concepts and new developments,''}
Kluwer Academic (Dordrecht, Netherlands, 1998).

\bibitem{kanPenrose}
R. Penrose, presented at the Cambridge University Seminar, {\it unpublished}
(1974).

\bibitem{kanCardoso1} V.~Cardoso and J.~P.~S.~Lemos,
{\it Phys. Lett.} {\bf B538}, 1 (2002); 
{\it Phys. Rev.} {\bf D67}, 084005 (2003).

\bibitem{kanCardoso3} U. Sperhake, V. Cardoso, F. Pretorius, E. Berti and J.A. Gonzalez,
arXiv:0806.1738 [gr-qc]. 


\bibitem{kanEG} D.~M.~Eardley and S.~B.~Giddings,
{\it Phys. Rev.} D {\bf 66}, 044011 (2002). 

\bibitem{kanYN} H.~Yoshino and Y.~Nambu,
{Phys. Rev.} D{\bf 66}, 065004 (2002); 
{ Phys.\ Rev.} D {\bf 67}, 024009 (2003). 

\bibitem{kanVoloshin}
M.~B.~Voloshin, 
{\it Phys. Lett.} {\bf B518}, 137 (2001); 
{\it Phys. Lett.} {\bf B524}, 376 (2002). 

\bibitem{kanDimEmp} S.~Dimopoulos and R.~Emparan,
{\it Phys.\ Lett.} {\bf B526}, 393 (2002).

\bibitem{kanKolVen}
E.~Kohlprath and G.~Veneziano,
{\it JHEP} {\bf 0206}, 057 (2002). 

\bibitem{kanSolodukhin} S.~N.~Solodukhin,
{\it Phys. Lett.} {\bf B533}, 153 (2002). 

\bibitem{kanRizzo} T.~G.~Rizzo,
{\it JHEP} {\bf 0202}, 011 (2002). 

\bibitem{kanHsu} S.~D.~H.~Hsu, 
{\it Phys. Lett.} {\bf B555}, 92 (2003). 

\bibitem{kanYR} H.~Yoshino and V.~S.~Rychkov,
{\it Phys. Rev.}  {\bf D71}, 104028 (2005). 

\bibitem{kanDas} M.~Cavaglia, S.~Das and R.~Maartens,
{\it Class. Quant. Grav.} {\bf 20}, L205 (2003).

\bibitem{kanYoshMann}
H.~Yoshino and R.~B.~Mann,
{\it Phys. Rev.}  {\bf D74}, 044003 (2006).


\bibitem{kanIOP1} D.~Ida, K.~y.~Oda and S.~C.~Park,
Phys.\ Rev.\  D {\bf 67}, 064025 (2003)
[Erratum-ibid.\  D {\bf 69}, 049901 (2004)]. 

\bibitem{kanKoch} B.~Koch,
{\it Phys. Lett.} {\bf B663/4}, 334 (2008).

\bibitem{kanGiddings} S.~B.~Giddings,
Phys.\ Rev.\  D {\bf 67}, 126001 (2003).

\bibitem{kanGT} S.~B.~Giddings and S.~Thomas,
{Phys. Rev.} D {\bf 65}, 056010 (2002). 

\bibitem{kanDL} S.~Dimopoulos and G.~Landsberg, 
{Phys. Rev.} D {\bf 87}, 161602 (2001).

\bibitem{kanADMR} P.~C.~Argyres, S.~Dimopoulos and J.~March-Russell,
{Phys. Lett.} B {\bf 441}, 96 (1998). 

\bibitem{kanreview} P.~Kanti,
Int.\ J.\ Mod.\ Phys.\  A {\bf 19}, 4899 (2004). 

\bibitem{kanHawking} S.~W.~Hawking, 
{\it Commun. Math. Phys.} {\bf 43}, 199 (1975). 

\bibitem{kanUnruh} W.~G.~Unruh,
Phys.\ Rev.\  D {\bf 10}, 3194 (1974); 
Phys.\ Rev.\  D {\bf 14}, 3251 (1976). 

\bibitem{kanOW} A.~C.~Ottewill and E.~Winstanley,
Phys.\ Rev.\  D {\bf 62}, 084018 (2000).

\bibitem{kanZeldovich} Y.B. Zel'dovich, {\it JETP Lett.} {\bf 14}, 180 (1971).
  
\bibitem{kanKMR1} P.~Kanti and J.~March-Russell,
{Phys. Rev.} D {\bf 66}, 024023 (2002). 

\bibitem{kanKMR2} P.~Kanti and J.~March-Russell,
{Phys. Rev.} D {\bf D67}, 104019 (2003). 

\bibitem{kanFS} V.~Frolov and D.~Stojkovic,
{Phys. Rev.} D {\bf 67}, 084004 (2003). 

\bibitem{kanHK2} C.~M.~Harris and P.~Kanti,
Phys.\ Lett.\  B {\bf 633}, 106 (2006). 

\bibitem{kanDHKW} G.~Duffy, C.~Harris, P.~Kanti and E.~Winstanley,
JHEP {\bf 0509}, 049 (2005). 

\bibitem{kanIOP2} D.~Ida, K.~y.~Oda and S.~C.~Park,
Phys.\ Rev.\  D {\bf 71}, 124039 (2005). 

\bibitem{kanCKW} M.~Casals, P.~Kanti and E.~Winstanley,
JHEP {\bf 0602}, 051 (2006). 

\bibitem{kanIOP3} D.~Ida, K.~y.~Oda and S.~C.~Park,
Phys.\ Rev.\  D {\bf 73}, 124022 (2006). 

\bibitem{kanCDKW} M.~Casals, S.~R.~Dolan, P.~Kanti and E.~Winstanley,
JHEP {\bf 0703}, 019 (2007).

\bibitem{kanNP} E.~Newman and R.~Penrose, 
{J. Math. Phys.} {\bf 3}, 566 (1962). 

\bibitem{kanChandra}
S.~Chandrasekhar, {\it The Mathematical Theory of Black Holes} (Oxford
University Press, New York, 1983).

\bibitem{kanGoldberg} J.~N.~Goldberg, A.~J.~MacFarlane, E.~T.~Newman,
F.~Rohrlich and E.~C.~Sudarshan, 
{J. Math. Phys.} {\bf 8}, 2155 (1967). 

\bibitem{kanTeukolsky}
S.~A.~Teukolsky, 
{Phys. Rev. Lett.} {\bf 29}, 1114 (1972); 
{Astrophys. J.} {\bf 185}, 635 (1973). 

\bibitem{kanGKT} S.~S.~Gubser, I.~R.~Klebanov and A.~A.~Tseytlin,
{\it Nucl. Phys.} {\bf B499}, 217 (1997). 

\bibitem{kanHK} C.~M.~Harris and P.~Kanti,
{JHEP} {\bf 0310}, 014 (2003). 

\bibitem{kanPage} D.~N.~Page,
{Phys. Rev.} D {\bf 13}, 198 (1976). 

\bibitem{kanFlammer} C.~Flammer, {\it Spheroidal Wave Functions} (Stanford
University Press, Stanford, USA, 1957).

\bibitem{kanStaro} A.~A.~Starobinskii and S.~M.~Churilov,
{Sov. Phys.-JETP} {\bf 38}, 1 (1974).

\bibitem{kanFackerell} E.~D.~Fackerell and R.~G.~Crossman,
{J. Math. Phys.} {\bf 18}, 1849 (1977). 

\bibitem{kanSeidel} E.~Seidel,
{Class. Quant. Grav.} {\bf 6}, 1057 (1989). 

\bibitem{kanCEKT2} S.~Creek, O.~Efthimiou, P.~Kanti and K.~Tamvakis,
Phys.\ Rev.\  D {\bf 75}, 084043 (2007).

\bibitem{kanCEKT3} S.~Creek, O.~Efthimiou, P.~Kanti and K.~Tamvakis,
Phys.\ Rev.\  D {\bf 76}, 104013 (2007). 

\bibitem{kanMuller} C.~Muller, in {\it Lecture Notes in Mathematics: Spherical Harmonics}
(Springer-Verlag, Berlin-Heidelberg, 1966).

\bibitem{kanEHM}  R.~Emparan, G.~T.~Horowitz and R.~C.~Myers,
{Phys. Rev. Lett.} {\bf 85}, 499 (2000). 

\bibitem{kanKI} H.~Kodama and A.~Ishibashi,
{Prog. Theor. Phys.} {\bf 110}, 701 (2003); 
{Prog. Theor. Phys.}  {\bf 111}, 29 (2004). 

\bibitem{kanCNS} A.~S.~Cornell, W.~Naylor and M.~Sasaki,
JHEP {\bf 0602}, 012 (2006).

\bibitem{kanCEKT1} S.~Creek, O.~Efthimiou, P.~Kanti and K.~Tamvakis,
Phys.\ Lett.\  B {\bf 635}, 39 (2006).

\bibitem{kanCCG} V.~Cardoso, M.~Cavaglia and L.~Gualtieri,
Phys.\ Rev.\ Lett.\  {\bf 96}, 071301 (2006)
[Erratum-ibid.\  {\bf 96}, 219902 (2006)];
JHEP {\bf 0602}, 021 (2006).

\bibitem{kanJP} D.~K.~Park,
Phys.\ Lett.\  B {\bf 638}, 246 (2006).

\bibitem{kan-BGK} J.~Grain, A.~Barrau and P.~Kanti,
Phys.\ Rev.\  D {\bf 72}, 104016 (2005). 

\bibitem{kanCCDN} H.~T.~Cho, A.~S.~Cornell, J.~Doukas and W.~Naylor,
Phys.\ Rev.\  D {\bf 77}, 016004 (2008).

\bibitem{kanCEKT4} S.~Creek, O.~Efthimiou, P.~Kanti and K.~Tamvakis,
Phys.\ Lett.\  B {\bf 656}, 102 (2007). 

\bibitem{kanCDKW2}  M.~Casals, S.~R.~Dolan, P.~Kanti and E.~Winstanley,
{\it JHEP} {\bf 0806}, 071 (2008).

\bibitem{kanHPPRSW} C.~M.~Harris, M.~J.~Palmer, M.~A.~Parker, P.~Richardson,
A.~Sabetfakhri and B.~R.~Webber,
JHEP {\bf 0505}, 053 (2005).

\bibitem{kan-Char} C.~M.~Harris, P.~Richardson and B.~R.~Webber,
JHEP {\bf 0308}, 033 (2003). 

\bibitem{kan-Cat} M.~Cavaglia, R.~Godang, L.~Cremaldi and D.~Summers,
Comput.\ Phys.\ Commun.\  {\bf 177}, 506 (2007). 

\bibitem{kan-True} G.~L.~Landsberg,
J.\ Phys.\ G {\bf 32}, R337 (2006).

\bibitem{kan-HER} G.~Corcella {\it et al.},
JHEP {\bf 0101}, 010 (2001). 

\bibitem{kanACN} Y.~Aharonov, A.~Casher and S.~Nussinov,
Phys.\ Lett.\  B {\bf 191}, 51 (1987).

\bibitem{kanBDDO} T.~Banks, A.~Dabholkar, M.~R.~Douglas and M.~O'Loughlin,
Phys.\ Rev.\  D {\bf 45}, 3607 (1992).

\bibitem{kanBO} T.~Banks and M.~O'Loughlin,
Phys.\ Rev.\  D {\bf 47}, 540 (1993).

\bibitem{kanBOS} T.~Banks, M.~O'Loughlin and A.~Strominger,
Phys.\ Rev.\  D {\bf 47}, 4476 (1993).

\bibitem{kanGidd} S.~B.~Giddings,
Phys.\ Rev.\  D {\bf 49}, 947 (1994). 

\bibitem{kanKBH} B.~Koch, M.~Bleicher and S.~Hossenfelder,
JHEP {\bf 0510}, 053 (2005). 

\bibitem{kanHKS} T.~J.~Humanic, B.~Koch and H.~Stoecker,
Int.\ J.\ Mod.\ Phys.\  E {\bf 16}, 841 (2007).

\bibitem{kanKBS} B.~Koch, M.~Bleicher and H.~Stoecker,
J.\ Phys.\ G {\bf 34}, S535 (2007). 

\bibitem{kanGidreview} S.~B.~Giddings,
AIP Conf.\ Proc.\  {\bf 957}, 69 (2007). 

\bibitem{kan-KGB} P.~Kanti, J.~Grain and A.~Barrau,
Phys.\ Rev.\  D {\bf 71}, 104002 (2005). 

\bibitem{kanHNS} T.~Harmark, J.~Natario and R.~Schiappa,
arXiv:0708.0017 [hep-th].

\bibitem{kan-HB} R.~Bousso and S.~W.~Hawking,
Phys.\ Rev.\  D {\bf 54}, 6312 (1996). 


\end{thebibliography}
\end{document}